\documentclass[3p,times]{elsarticle}

\usepackage{ecrc}

\usepackage{tabularx,ragged2e}
\usepackage{multirow}
\usepackage{booktabs}
\usepackage{upgreek} 
\usepackage{lscape}
\usepackage{graphicx,verbatimbox}
\usepackage{listings}
\usepackage{subcaption}
\usepackage{mathtools}
\usepackage{color}
\usepackage{tabularx,ragged2e}
\usepackage{mathtools}
\usepackage{enumitem}
\usepackage[british]{babel}
\usepackage{xspace} 

\newcommand{\mt}{MeSH terms\xspace}
\newcommand{\mts}{MeSH term suggestion\xspace}
\newcommand{\fgm}{fragment\xspace}
\newcommand{\fgms}{fragments\xspace}
\newcommand{\ftac}{free text atomic clause\xspace}
\newcommand{\ftacs}{free text atomic clauses\xspace}
\newcommand{\bt}{BERT\xspace}
\newcommand{\atb}{Atomic BERT\xspace}
\newcommand{\seb}{Semantic BERT\xspace}
\newcommand{\frb}{Fragment BERT\xspace}

\volume{00}

\firstpage{1}

\journalname{Intelligent Systems with Applications}

\runauth{}

\jid{procs}

\jnltitlelogo{ISWA}

\usepackage{amssymb}
\usepackage{amsthm}

\usepackage[figuresright]{rotating}
\usepackage[textsize=tiny]{todonotes}
\begin{document}

\begin{frontmatter}



\dochead{}
 
\title{Automated MeSH Term Suggestion for Effective Query Formulation in Systematic Reviews Literature Search\footnote{This paper is currently in submission with Intelligent Systems with Applications Journal Technology-Assisted Review Systems Special issue and is under peer review.}}
 \cortext[cor1]{Corresponding Author}
 
 \author{Shuai Wang\corref{cor1}}
 \address{The University of Queensland, Brisbane, Australia}
 \ead{shuai.wang2@uq.edu.au}
 
 \author{Harrisen Scells}
  \address{The University of Queensland, Brisbane, Australia}
  \ead{h.scells@uq.edu.au}
  
    \author{Bevan Koopman}
     \address{CSIRO, Brisbane, Australia}
  \ead{bevan.koopman@csiro.au}
  
 \author{Guido Zuccon}
      \address{The University of Queensland, Brisbane, Australia}
 \ead{g.zuccon@uq.edu.au}



\address{}

\begin{abstract}
	High-quality medical systematic reviews require comprehensive literature searches to ensure the recommendations and outcomes are sufficiently reliable. Indeed, searching for relevant medical literature is a key phase in constructing systematic reviews and often involves domain (medical researchers) and search (information specialists) experts in developing the search queries. Queries in this context are highly complex, based on Boolean logic, include free-text terms and index terms from standardised terminologies (e.g., the Medical Subject Headings (MeSH) thesaurus), and are difficult and time-consuming to build. The use of MeSH terms, in particular, has been shown to improve the quality of the search results. However, identifying the correct MeSH terms to include in a query is difficult: information experts are often unfamiliar with the MeSH database and unsure about the appropriateness of MeSH terms for a query. Naturally, the full value of the MeSH terminology is often not fully exploited.
	
	This article investigates methods to suggest MeSH terms based on an initial Boolean query that includes only free-text terms. In this context, we devise lexical and pre-trained language models based methods. These methods promise to automatically identify highly effective MeSH terms for inclusion in a systematic review query. Our study contributes an empirical evaluation of several MeSH term suggestion methods. We further contribute an extensive analysis of MeSH term suggestions for each method and how these suggestions impact the effectiveness of Boolean queries.
\end{abstract}


\end{frontmatter}


\section{Introduction}

A medical systematic review is a comprehensive review of literature for a highly focused research question. Systematic reviews are seen as the highest form of evidence and are used extensively in healthcare decision making and clinical medical practice. In order to synthesise literature into a systematic review, a search must be undertaken. A major component of this search is a Boolean query. The Boolean query is often developed by a trained expert (i.e., an information specialist), who works closely with the research team to develop the search, and usually has some knowledge of the domain been searched. 
The most commonly used database for searching medical literature is PubMed. Due to the increasing size and scope of these databases, and of PubMed in particular, the Medical Subject Headings (MeSH) thesaurus was developed to conceptually index studies~\cite{zieman1997conceptual, richter2012using}. MeSH is a controlled vocabulary thesaurus arranged in a hierarchical tree structure (specificity increases with depth in a parent$\rightarrow$child relationship, e.g., \texttt{Anatomy}$\rightarrow$\texttt{Body Regions}$\rightarrow$\texttt{Head}$\rightarrow$\texttt{Eye}\dots etc.). Indexing and categorising studies with MeSH terms enables queries to be developed which incorporate both free-text keywords \textit{and} MeSH terms --- enabling more effective searches.
The use of MeSH terms in queries has been shown to be more effective than free-text keywords alone~\cite{richter2012using, chang2006searching, abdou2008searching, tenopir1985full}, e.g., they increase precision~\cite{liu2017evaluating} and are far less ambiguous than free-text~\cite{wacholder1997disambiguation}. 
However, it is still difficult even for expert information specialists to be familiar with the entire MeSH controlled vocabulary~\cite{liu2009impact, liu2017evaluating} --- at the time of writing, MeSH contains 29,640 unique headings.

PubMed has attempted to overcome this difficulty by developing a method called Automatic Term Mapping (ATM). 
ATM is an automatic query expansion method which attempts to seamlessly map free-text keywords in a query to one of the three categories (index tables): MeSH, journal name or author name~\cite{nahin2003change}.
Although ATM is applied by default for all queries issued to PubMed, it has several semantic limitations: it is inaccurate when used to expand free-text acronyms into MeSH terms~\cite{schulz2001indexing}; it produces different MeSH expansions even though synonymic free-text terms are used~\cite{adlassnig2009optimization}; and has difficulty disambiguating between MeSH terms and journal names~\cite{smith2004examination}. Despite these limitations, the use of ATM for MeSH term suggestion has been shown to increase the precision of free-text searches in the genomic domain~\cite{lu2009evaluation}, and is the state-of-the-art method for the MeSH Term suggestion task. However, its use has, to the best of the authors' knowledge, not been empirically evaluated in the context of improving the effectiveness of systematic review literature search queries.

Recent advances in the use of pre-trained language models (PLMs) such as BERT \cite{DBLP:conf/naacl/DevlinCLT19}, T5 \cite{raffel2019exploring}, and GPT-3 \cite{brown2020language} have delivered state-of-the-art performance in many natural language processing tasks. Typically, a pre-trained language model is trained on a large corpus using the transformer architecture to ``get familiar'' with language representations. Then the model is fine-tuned to downstream tasks to perform with high effectiveness across the target task. The transformer architecture is an encoder-decoder model training structure that does not use recurrence and convolutions~\cite{vaswani2017attention}. Prior work showed that using PLMs can significantly increase effectiveness in ad-hoc search~\cite{lin2021pretrained} as well as in professional search~\cite{Eugene2022ecirtar, chalkidis-etal-2020-legal,QIN2021121,choe2022short}.


In this article, we introduce the task of MeSH term suggestion for Boolean queries used in systematic review literature search \footnote{This article is an extension of our previous work published at the 2021 Australasian Document Computing Symposium~\cite{wang2021mesh}.}. We model this task within the context of an information specialist looking for MeSH terms to add to a query without MeSH terms currently present. We also propose a framework to evaluate the effectiveness of the suggestion of MeSH terms on established collections of systematic review literature search queries. This article adds to a recent stream of research that has focused on computational methods for the assisted formulation~\cite{scells2020automatic,scells2020objective,scells2021comparison, agosti2019analysis} or refinement~\cite{wang2021mesh, scells2018generating, scells2019www, agosti2020post, alharbi2020refining} of Boolean queries for systematic review creation, and more generally to research on computational methods for technology-assisted reviews~\cite{li2020stop, sneyd2021stopping, lee2022towards, lee2018seed, cormack2017technology}. Furthermore, we propose two categories of methods for the MeSH term suggestion task, including methods based on the BERT pre-trained language model and methods not based on BERT (lexical methods). We show that our methods suggest MeSH terms that outperform the effectiveness of the MeSH terms selected by the information specialists and included in the original queries. Our methods are readily integrable into tools for information specialists to help with the construction of systematic review Boolean queries.

The contributions of this article are:

\begin{enumerate}
	\item The introduction of the new task of suggesting MeSH terms for systematic review literature search (Boolean queries), modelled within the context of an information specialist looking for MeSH terms to add to a query without MeSH terms present.
\item The formulation of MeSH term suggestion methods to help information specialists and researchers to construct Boolean queries for systematic review creation.

\item An empirical evaluation of the effectiveness of different MeSH terms suggestion methods

\item An understanding of how the MeSH terms suggested by the proposed automatic methods differ from those originally selected by information specialists formulating the query.
	

\end{enumerate}

\section{Material and methods}

\subsection{Overview of the MeSH Term Suggestion Task}
\begin{figure}[t!]
 	\centering
 	\includegraphics[width=0.9\textwidth]{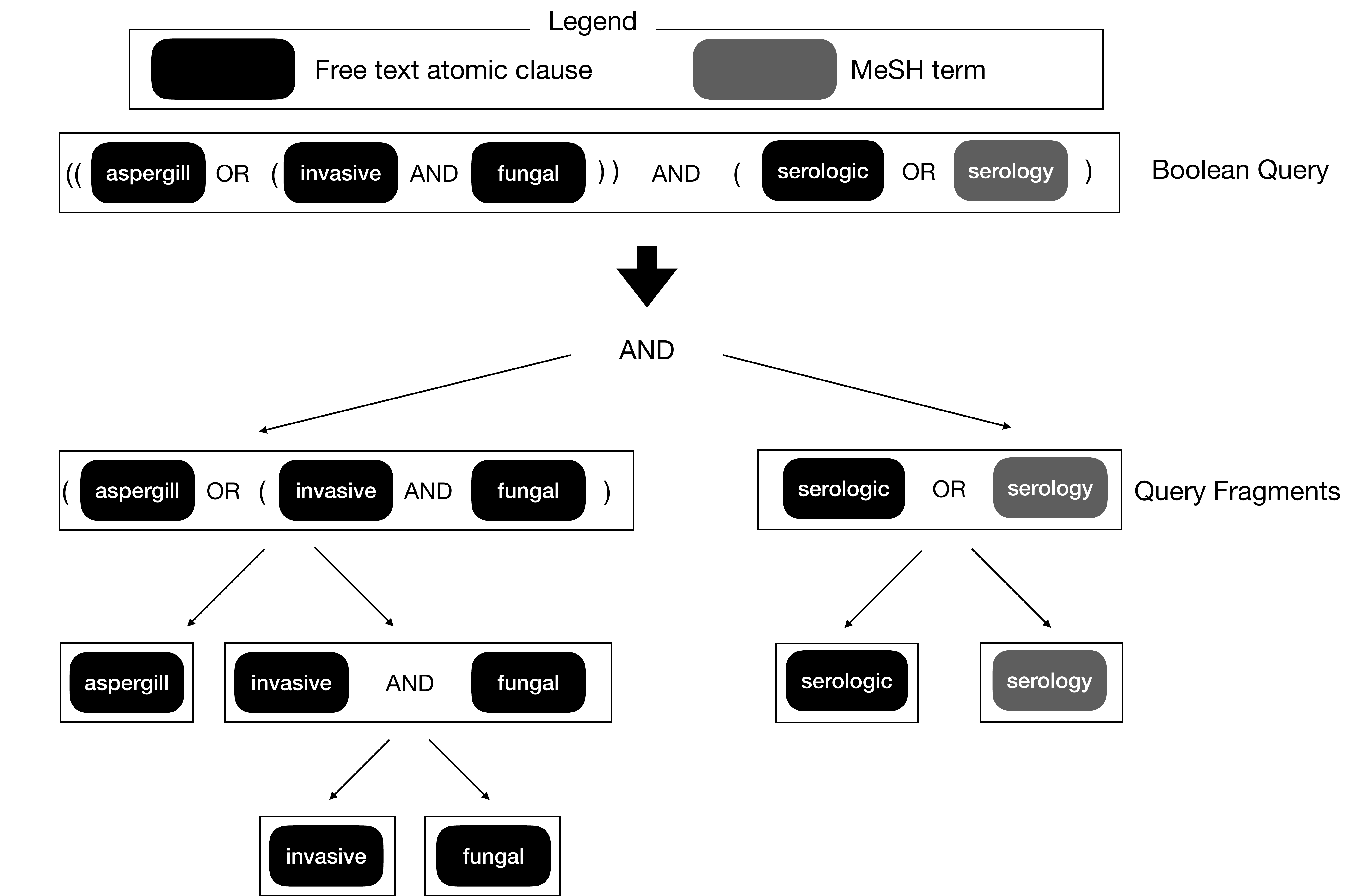}
 	\caption{Example query showing a \textbf{Boolean Query}, two \textbf{Query Fragments}, several \textbf{Free text atomic clauses}, and a \textbf{MeSH term}.}
 	\label{fig:query-concept}
 \end{figure}

We start by outlining the task of MeSH term suggestion for Boolean queries that do not already contain MeSH terms.
We assume the user has entered a Boolean query without MeSH terms. A Boolean query can be viewed as a tree where Boolean operators (e.g. AND, OR) represent the internal nodes of the tree, while free-text atomic clauses and MeSH Terms are the leaves. Free-text atomic clauses are one or more words that express a concept, e.g., a disease, a treatment or a population aspect. We call each of the first level nodes of the tree (i.e. the nodes at depth 1) a query fragment. Typically, a query fragment represents an individual aspect of an information need~\cite{suhail2013methods}; specifically, each query fragment corresponds to a different PICO element, i.e. population, intervention, control, and outcome~\cite{schardt2007utilization}.  
These concepts are shown in Figure~\ref{fig:query-concept}. The task of MeSH term suggestion is to identify appropriate MeSH terms to be added as leaves to a query fragment. 
In this article, we suggest MeSH terms for each query fragment independently of each other. We leave the investigation of query fragment dependencies concerning MeSH term suggestions for future work.
 
\begin{figure}[t!]
	\centering
	\includegraphics[width=\textwidth]{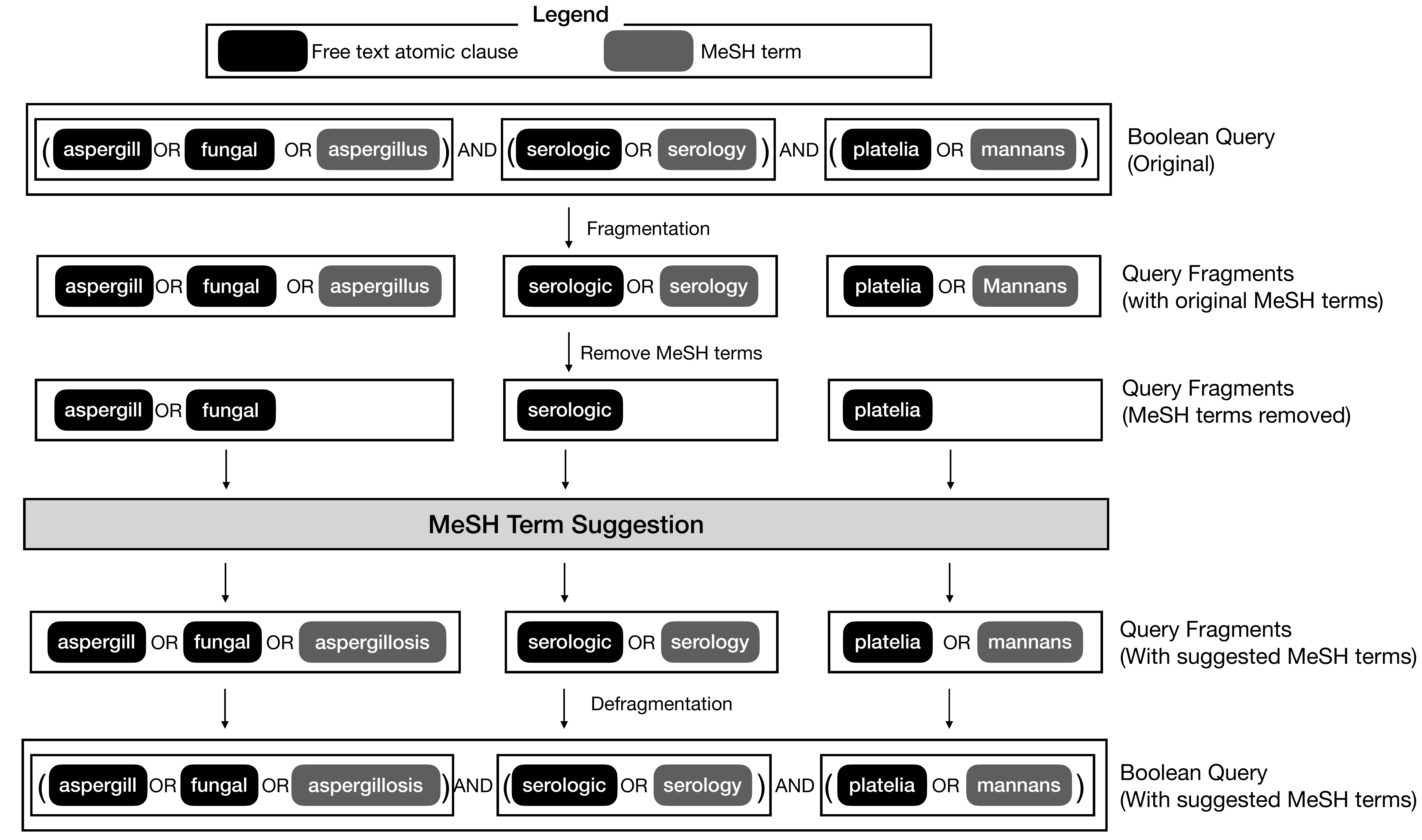}
	\caption{Overview of the MeSH term suggestion procedure. Proposed methods using lexical MeSH term retrieval or BERT MeSH term retrieval facilitate the suggestion of MeSH terms. We evaluate each method that suggests MeSH terms in terms of (1) the ability for the suggested MeSH terms to effectively retrieve literature for a defragmented Boolean query , (2) overlap between suggested MeSH terms and MeSH terms included in the original query. Note that the number of MeSH terms suggested for a fragment may be lower or higher than the number of MeSH terms in the original query.}
	\label{fig:query-overview}
\end{figure}

Figure~\ref{fig:query-overview} gives an intuition for how we obtain query fragments from a Boolean query, how MeSH terms are suggested for a given query fragment, and how we perform \textit{defragmentation} to construct a new Boolean query that includes MeSH terms. 
%
The figure shows that after fragmentation (i.e. the process of deriving query fragments), we remove all the MeSH terms from each query fragment. We then apply a MeSH term suggestion technique which adds new MeSH terms into a query fragment. The new query fragments that now contain suggested MeSH terms are then defragmented by combining all of the query fragments corresponding to the original query with the \texttt{AND} operator.

This work extends our existing line of research into MeSH term suggestion \cite{wang2021mesh}, where we previously developed several techniques that depend on pre-existing lexical matching systems. One limitation of these systems is their dependence on manually crafted rules that are expensive to create and have limitations in terms of how words are matched to MeSH terms (e.g., spelling variants, acronyms, misspellings).
This article instead investigates the use of pre-trained language models, i.e., BERT, for the task of MeSH term suggestion. These neural models have been shown to be resilient to the shortcomings of lexical-based systems~\cite{DBLP:conf/naacl/DevlinCLT19, wang2021bert}. However, neural models have their own limitations, particularly requiring large amounts of training data. 
The following sections first provide a brief overview of our existing lexical-based techniques and then describe our new neural techniques in detail, specifically addressing the need for ad-hoc training data.

\subsection{Lexical MeSH Term Suggestion}

Our lexical-based methods are formulated as a pipeline of three steps: retrieval, ranking, and refinement. The following sections provide a brief overview of each of these steps. For a more comprehensive discussion of the lexical-based methods, refer to our previous work~\cite{wang2021mesh}.

\paragraph{Retrieval}
\label{section:mesh_retrieval}

The first step in our \mts pipeline is the \textbf{retrieval} of \mt. The retrieval of \mt is facilitated by three different methods:

\begin{description}
	\item[ATM] The entire free-text only query fragment is submitted to the PubMed Entrez API~\cite{sayers2010general} for \textit{automatic term mapping} (ATM). This is the default system used by PubMed for automatically adding MeSH terms to queries. 
	\item[MetaMap] Each free-text atomic clause in a query fragment is submitted to MetaMap~\cite{aronson2001effective}.\footnote{Version 2018 with options set to default values.} The results are filtered to only include those entities derived from the MeSH source. All of the mapped \mt are recorded for each of the free-text terms in a query fragment. Additionally, the score is recorded for each MeSH term.
	\item[UMLS] We index UMLS~\cite{bodenreider2004unified}\footnote{version 2019AB using the \texttt{MRCONSO}, \texttt{MRDEF}, \texttt{MRREL}, and \texttt{MRSTY} tables.} into Elasticsearch v7.6. Each free-text atomic clause in the query fragment with MeSH terms removed is submitted to the Elasticsearch index. The results are filtered to only include synonyms of concepts derived from the MeSH source. Additionally, the BM25 score is recorded for each MeSH term.
\end{description}

For the MetaMap and UMLS approaches, the same MeSH term may be retrieved multiple times for a given free-text fragment. To overcome this issue, we re-score the \mt using rank fusion (CombSUM)~\cite{fox1994combination}. The intuition for this re-scoring is that highly common \mt that also obtain a high score from these retrieval methods should be scored highly overall (thus ranked higher than common \mt \textit{and} highly scoring \mt).

\paragraph{Ranking}
\label{section:mesh_ranking}

Once MeSH terms have been retrieved, they are ranked according to the approach for entity ranking described by Jimmy et al.~\cite{jimmy2019health} by adapting features proposed by Balog~\cite{balog2018entity}. In total, we use eleven entity features. 
Positive instances correspond to \mt in the original query fragment; negative instances correspond to \mt not in the original query fragment (binary labels). 
With features and instance labels, we train a learning-to-rank (LTR) model for each retrieval method.
In addition to LTR, we also investigate a rank fusion approach~\cite{fox1994combination}, where we combine the normalized MeSH term suggestion scores from each of the three methods to produce a new ranking that incorporates the highest ranking MeSH terms from each method. The intuition for investigating rank fusion in this context is that each method may retrieve different MeSH terms; and those terms may be ranked differently each time. Therefore, we boost MeSH terms that are retrieved and ranked highly by multiple methods.

\paragraph{Refinement}
\label{section:mesh_refinement}
Finally, we seek to refine the suggested MeSH terms by estimating a rank cut-off. 
We do this using a score-based gain function. Formally, the cumulative gain $CG$ for a MeSH term at rank $p$ is 
\begin{equation}
	CG_p = \sum_{i=1}^{p}score_i
\end{equation}


\noindent where the score for a MeSH term is equal to $1-normalised\ score$ (i.e., min-max normalisation) for the MeSH term. 

We tune a parameter, $\kappa$, for each retrieval method which controls the percentage of total $CG$ allowed to be observed before the ranking is cut-off (i.e., a refinement of the ranking). We tune $\kappa$ from 5\% to 95\% in increments of 5\%.
The intuition for re-scoring MeSH terms becomes apparent when used with the $\kappa$ parameter: the highest-ranking MeSH term will receive a score of 0, resulting in at least one MeSH term suggested for every query fragment. 

Note that MeSH terms may share the same score, i.e., they may be tied. We take a conservative approach to account for the problem of tied MeSH terms at the boundary of the cut-off specified by $\kappa$. Whenever we encounter ties, we treat all of the tied MeSH terms as a single accumulation of gain that equals the summed gain across the scores of the tied MeSH terms. This treatment has the effect that tied MeSH terms account for much larger accumulations of gain. Therefore, tied MeSH terms at the top of rankings are more likely to be included in the cut-off than tied MeSH terms at the bottom. In essence, either all tied MeSH terms are considered within the cut-off (i.e., ties at the top of the ranking), or no tied MeSH terms are considered (i.e., ties at the bottom of the ranking). 

\subsection{BERT MeSH Term Suggestion}
\label{sec:BERT_suggester}
\begin{figure}[t!]
	\centering
	\includegraphics[width=\textwidth]{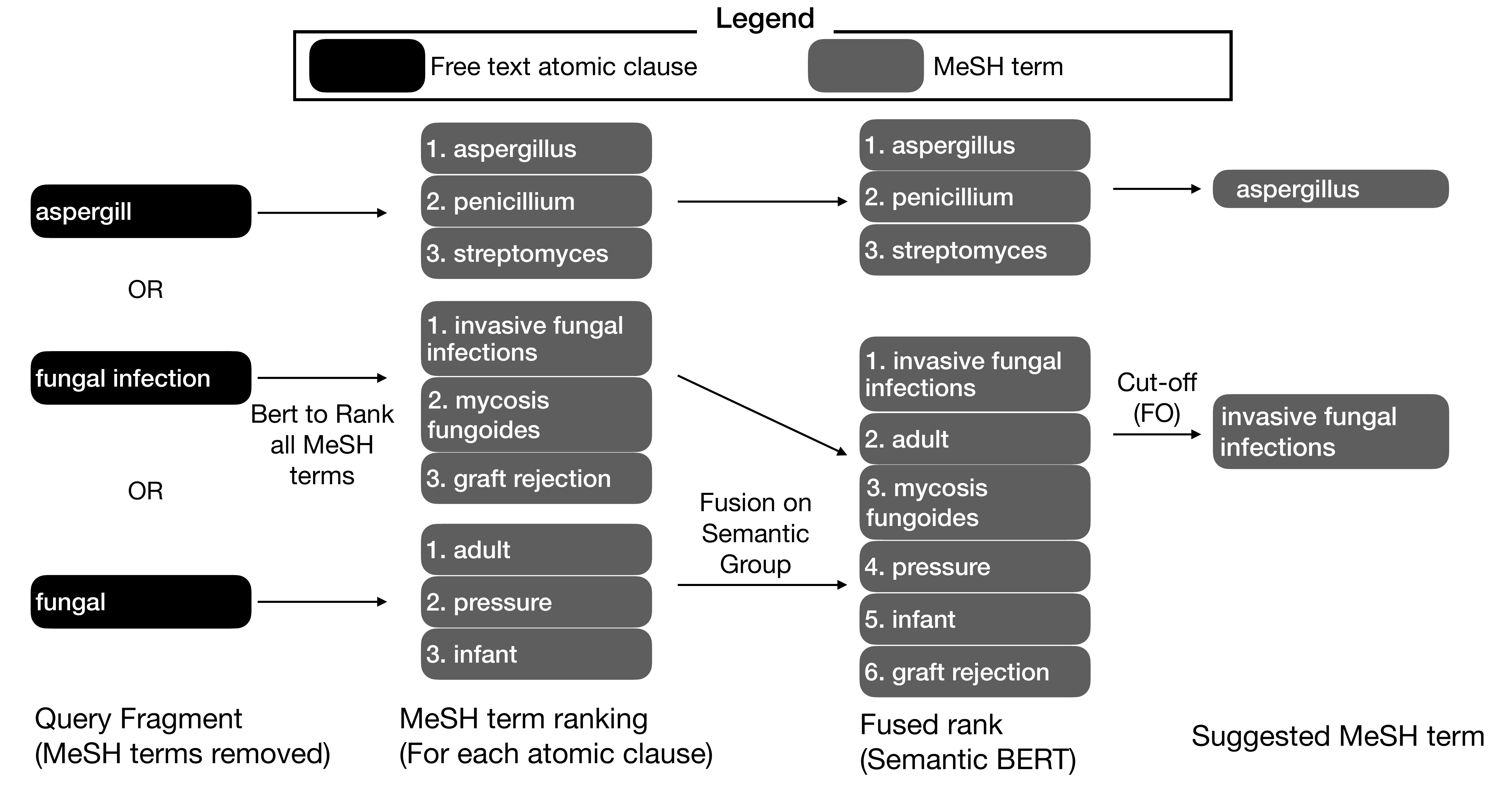}
	\caption{Overview of the MeSH term suggestion for the BERT methods. Note that Fusion of MeSH ranks may be optional in the pipeline.}
	\label{fig:BERT_suggestion}
\end{figure}

Next, we extend our MeSH term suggestion methods using fine-tuned PLM models. Firstly, PLM models are typically chosen from the same domain in which the task is conducted. 

\begin{figure}[t!]
	\centering
	\includegraphics[width=\textwidth]{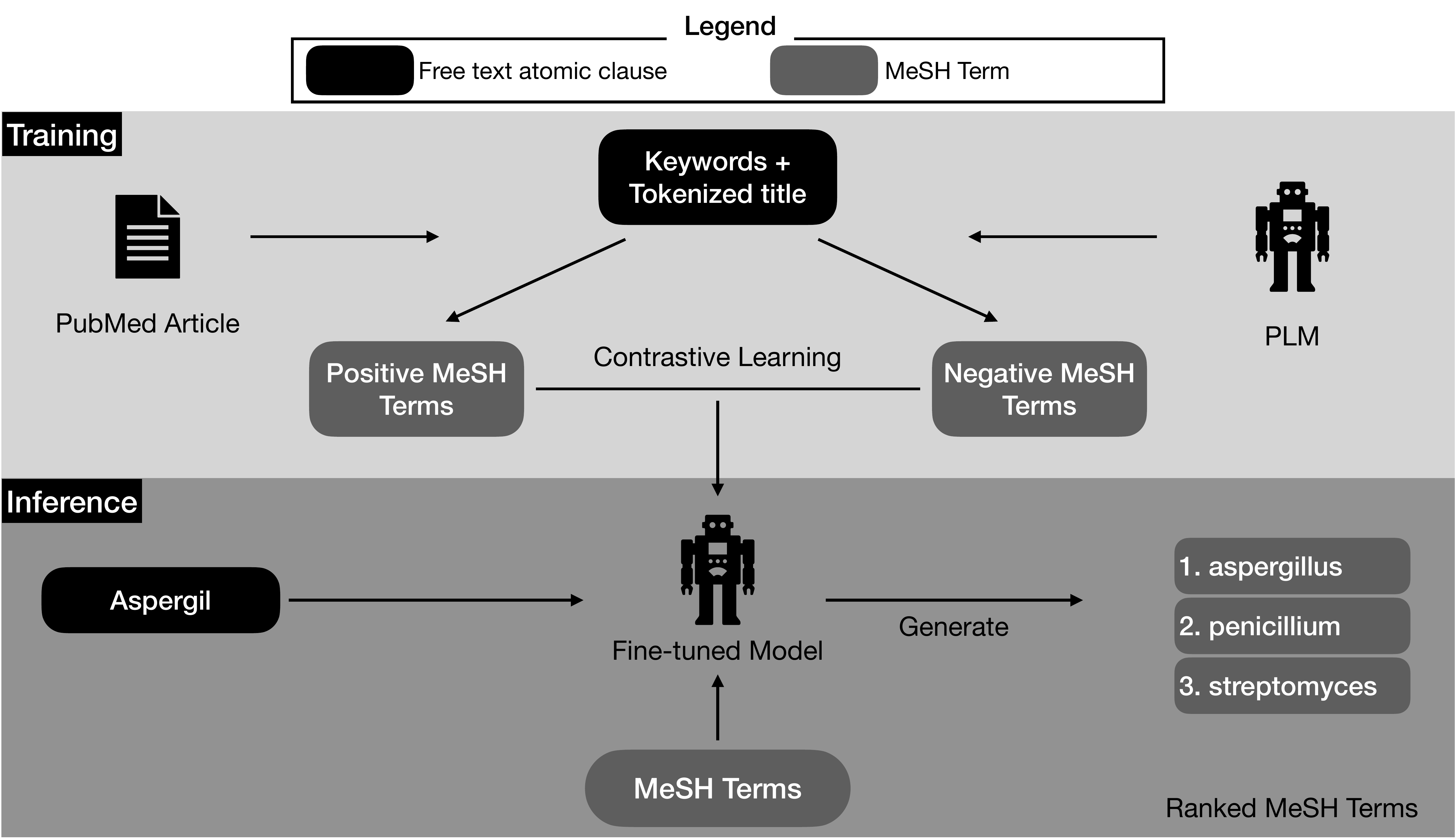}
	\caption{Architecture of model fine-tuning and inference.}
	\label{fig:architecture}
\end{figure}

\paragraph{Architecture}
We show the architecture of our fine-tuning and inference processes in Figure \ref{fig:architecture}. 
We use BioBERT~\cite{lee2020biobert} as the base PLM, as the context of this paper is medical systematic reviews. 
BioBERT is a PLM pre-trained on PubMed abstracts and PubMed Central (PMC)\footnote{PubMed Central is the repository containing full-text articles of the open-access part of the PubMed database.} full-text articles using the \bt training architecture \cite{DBLP:conf/naacl/DevlinCLT19}. After fine-tuning, BioBERT has achieved state-of-the-art performance on many medical-related tasks, including biomedical named entity recognition, relation extraction and question answering \cite{lee2020biobert}.

Ideally, training data closely related to the target task should be used to fine-tune a PLM to achieve the highest effectiveness. Ideally, in our case, we would use professionally constructed medical systematic review Boolean queries to fine-tune our model. However, PLMs are typically data-hungry and require a large number of labelled training samples. In systematic review literature search, several public datasets are available with Boolean queries, such as the CLEF TAR collections~\cite{kanoulas2017clef,kanoulas2018clef,kanoulas2019clef}, the collection of~\citet{wang2022little}, and the collection of \citet{scells2017test}. Between these datasets, however, only 253 unique topics would be available to train the model: an insufficient amount to effectively fine-tune a BERT model.

Instead, we create training samples by approximating the target task using data obtained from PubMed. We use the publicly available PubMed baseline to obtain the metadata about all published articles up to the start of 2022. 
The metadata contains information such as the title and abstract, but importantly for this work, it also includes author-assigned keywords and the relevant MeSH terms for an article.
We use the assigned keywords and MeSH terms for every article in the PubMed dataset to approximate the task of MeSH term suggestion. To maximise the amount of training data, we also extract keywords from the title (as not all PubMed articles contain keywords).
To tokenise titles, we use the process described by \citet{wang2022seed}. Firstly, we tokenise the title using Gensim~\cite{khosrovian2008gensim}, and then we remove stopwords too using NLTK~\cite{bird2004nltk}. 
We use the toolkit proposed by~\citet{Gao2022TevatronAE} to develop a dense retriever to suggest MeSH Terms.
The model is fine-tuned with localized contrastive loss using triples of $<k_{a,i},m_a^+,m_a^->$ where $a$ is a PubMed article, $k_{a,i}$ is the $i$th keyword in the PubMed article, $m_a^+$ are the MeSH terms for the PubMed article, and $m_a^-$ are ten randomly sampled MeSH terms from the MeSH thesaurus. Many MeSH terms contain spaces or punctuation. Our model considers each MeSH term a unique token in the model vocabulary.
Once the model is fine-tuned, we obtain an encoding for all MeSH terms. At inference time, we create an encoding for a keyword to obtain a score using the \texttt{[CLS]} token for all MeSH terms. Thus, our method scores and ranks all MeSH terms given a keyword.

\begin{table*}
	\centering
	\small
	\begin{tabular}{c|p{75pt}|p{75pt}|p{75pt}|p{75pt}}
		\toprule
		
		MeSH Removed Fragment& \multicolumn{4}{c}{neonatal sepsis OR neonatal bacteremia OR neonatal infections OR death} \\ \midrule
		\ftacs&neonatal sepsis & neonatal bacteremia& neonatal infections&death \\\midrule
		semantic group&\multicolumn{3}{c|}{neonatal sepsis, neonatal bacteremia, neonatal infections}&death \\
		\bottomrule
		
	\end{tabular}
	\caption{Example query fragments with separation of semantic groups. In the example, `neonatal sepsis', `neonatal bacteremia' and `neonatal infections' are grouped to form a semantic group, while `death' is another semantic group. }	
	\label{table:semantic_example}
\end{table*}

\paragraph{Ranking Suggestions} The goal of MeSH term suggestion is to suggest MeSH terms for each query fragment. However, the result from the BERT suggestion method consists of a ranked list of \mts for each \ftac. We need to combine the rankings for each MeSH term. We formulate this combination task into two steps, (1) choosing how we represent a MeSH term ranking, and (2) choosing where to cut off the ranking. We present an overview of the combination task in Figure \ref{fig:BERT_suggestion}.


First, we choose the best way to represent a ranking, which means deciding if MeSH terms should be suggested individually for every \ftacs, as a whole for every \fgm, or using other heuristics to decide how the representation should be computed. We designed three ranking representation methods:

\begin{enumerate}	
\item \textbf{\atb}: Firstly, we treat suggestions for each \ftac individually, essentially applying no strategy to combine the suggestions. 

\item \textbf{\frb}: Next, we study the combination of all MeSH term rankings for a given query fragment. We apply rank fusion (normalised CombSUM~\cite{fox1994combination}) to all of the \ftacs in a query fragment. For computational reasons, we only use the top 20 MeSH terms for each \ftac.

\item \textbf{\seb}: Finally, we study semantically grouping \ftacs and apply the same rank fusion technique as above, but this time to each group.
We show an example of a semantic group in Table \ref{table:semantic_example}. To derive semantic groups, we first take all \ftacs from the \fgm and obtain word2vec embeddings for each \ftac. Then we compute cosine similarities between all \ftac to decide if they are semantically related. In our experiments, we apply a threshold of 0.7 on the similarity.
We use a word2vec model pre-trained on PubMed and Wikipedia~\cite{moen2013distributional}. There are two reasons we use word2vec rather than BERT for semantic groups. First, if we apply our proposed BERT model, we note that we fine-tuned using semantic pairs of \ftacs and MeSH terms: thus, calculating the similarity between two \ftacs can cause a model mismatch. Secondly, the use of an additional BERT model will increase the latency in producing suggestions at inference time, as each \ftac needs to be encoded twice.

\end{enumerate}
Second, we choose where to cut off the ranking of MeSH terms from the ranking representations. We propose four strategies to cut off MeSH term rankings:
\begin{enumerate}
	\item \textbf{First only (FO)}: The first MeSH term of the ranking is selected for each ranking representation.
	\item \textbf{Same as \ftacs (SA)}: The number of MeSH terms selected equals the number of \ftacs in each \fgm (i.e, only applicable to \textbf{Fragment BERT}).
	\item \textbf{Same as original (SO)}: The MeSH terms selected equals the number of MeSH terms in the query fragment prior to removing MeSH terms (i.e, only applicable to \textbf{Fragment BERT}).
	\item \textbf{Linear (LN)}: The number of MeSH terms selected is learnt using a linear function with respect to the number of \ftacs in the fragments (i.e, only applicable to \textbf{Fragment BERT}).
\end{enumerate}


\subsection{Evaluation}
\label{methods:evaluation}

The end goal of a systematic review literature search is to find all of the relevant literature at the minimum cost. Thus, an effective Boolean query minimises the number of documents retrieved while maximising the retrieval of relevant documents. In our MeSH term suggestion task, we use the retrieval effectiveness of defragmented Boolean queries to evaluate MeSH term suggestion.

The MeSH terms included in the original query have been derived often after careful consideration by expert information specialists. We therefore consider how the MeSH terms included in the original queries differ from those suggested by the methods investigated in this work; specifically, we measure the overlap between the suggested MeSH terms and the MeSH terms included in the original query. We note that a MeSH term that is not in the original query may not necessarily be less effective of a search term than one included in the original query. 




To evaluate the effectiveness of the suggested MeSH terms for the task of systematic review literature search, once query fragments are defragmented, the retrieval effectiveness is evaluated using typical systematic review literature search evaluation measures: precision, recall, and ${F\beta}$, with $\beta=\{1,3\}$. The PubMed Entrez API is used to directly issue defragmented Boolean queries to obtain retrieval results. As PubMed is constantly updated with new studies, we apply a date restriction to all queries for reproducibility purposes. We use the Jaccard index measure to evaluate the overlap of MeSH terms between those suggested by the investigated methods and those included in the original query.



For both evaluation settings (i.e., Boolean query retrieval and evaluation against original MeSH terms), we evaluate the lexical suggestion method in two settings: (i) \textbf{all}, where all retrieved MeSH terms are considered; and (i) \textbf{cut}, where the score-based cut-off is used. We also evaluate all \bt suggestion methods and compare their effectiveness with that of the original query and the lexical methods. 

\subsection{Experimental Setup}

\begin{figure}[!t]
	\begin{minipage}{0.245\textwidth}
		\centering
		\includegraphics[width=1\linewidth]{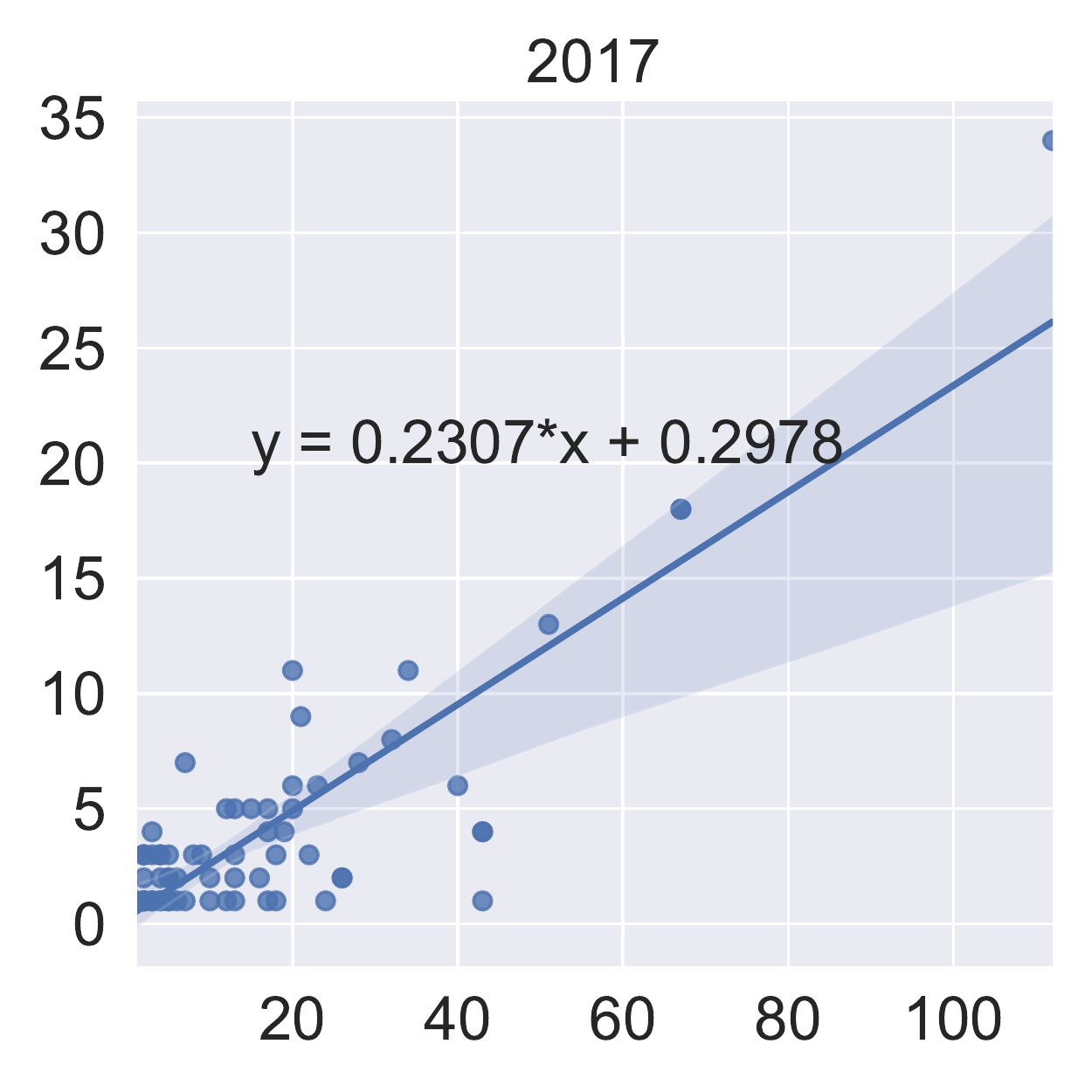}
	\end{minipage}\hfill
	\begin{minipage}{0.245\textwidth}
		\centering
		\includegraphics[width=1\linewidth]{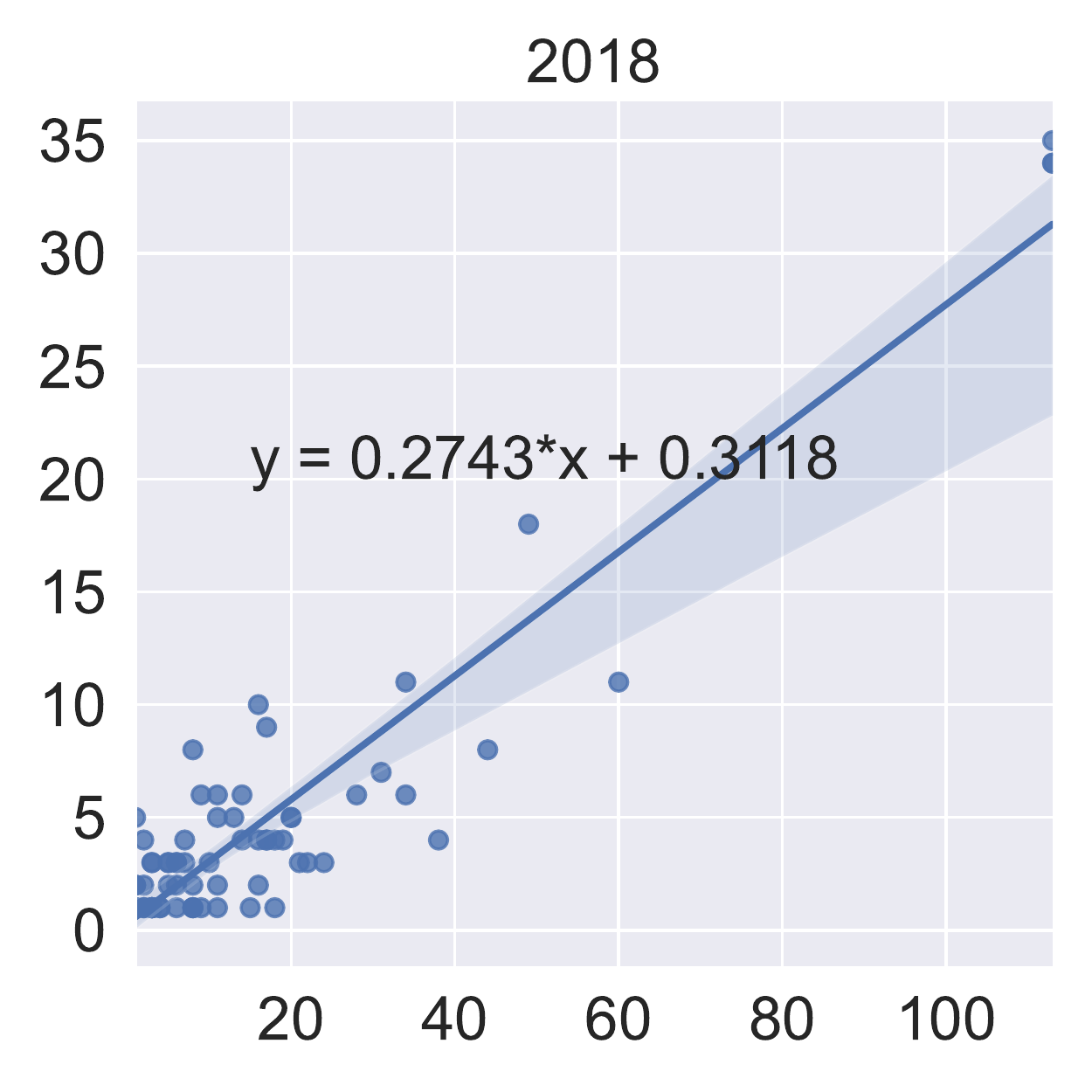}
	\end{minipage}\hfill
	\begin{minipage}{0.245\textwidth}
		\centering
		\includegraphics[width=1\linewidth]{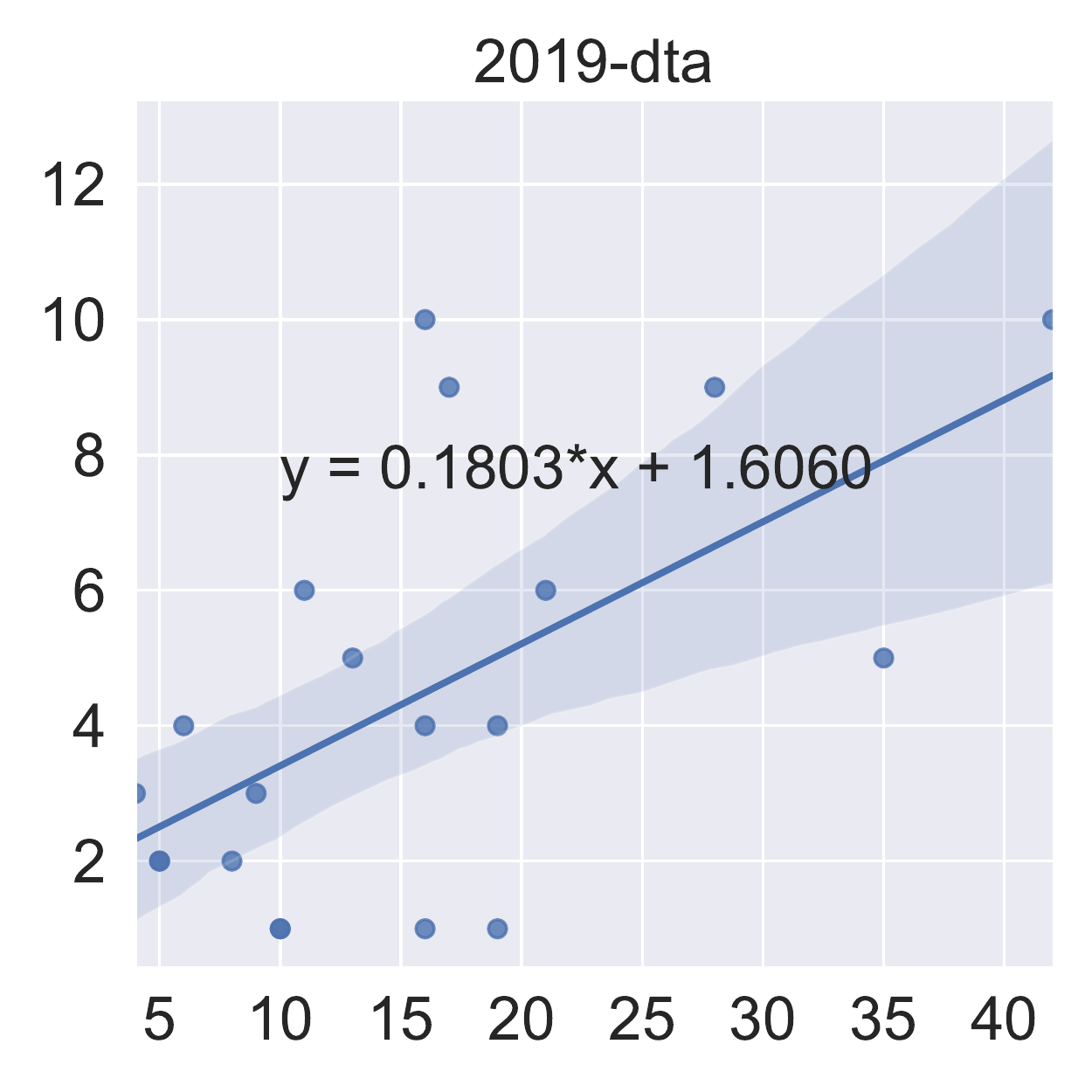}
	\end{minipage}\hfill
	\begin{minipage}{0.245\textwidth}
		\centering
		\includegraphics[width=1\linewidth]{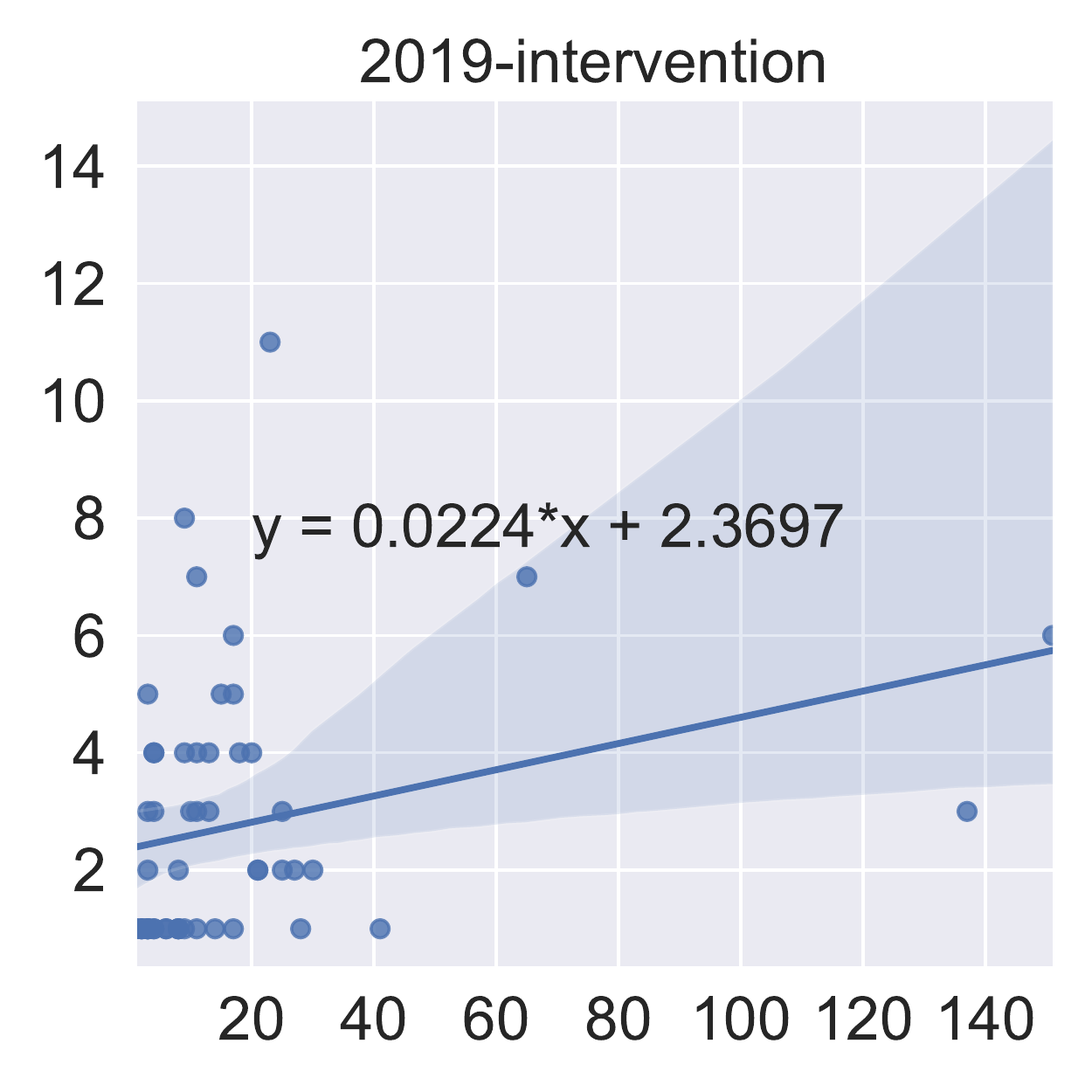}
	\end{minipage}
	\caption{Linear regression performed on the number of keywords (x-axis) and the number of MeSH terms (y-axis) in query fragments for training splits of CLEF TAR 2017, 2018, 2019-dta and 2019-intervention.} 
	\label{fig:learn_J}
	
\end{figure}

For our experiments, we use topics from the CLEF TAR task from 2017, 2018, and 2019 \cite{kanoulas2017clef, kanoulas2018clef, kanoulas2019clef}. 15 topics are discarded due to lack of MeSH terms\footnote{Discarded topics are: \textbf{2017}: CD007427, CD010771, CD010772, CD010775, CD010783, CD010860, CD011145; \textbf{2018}: CD007427, CD009263, CD009694; \textbf{2019}: CD006715, CD007427, CD009263, CD009694, CD011768.}. An additional topic is discarded because of retrieval issues\footnote{The additional discard topic is \textbf{2017}: CD010276.}, likely resulting from the fact that we translate queries automatically from one format (Ovid Medline) into another format (PubMed)~\cite{scells2018querylab}. In total, we used 116 unique topics, as each year has partial overlap. For each topic, we automatically divide the Boolean query for that topic into query fragments~\cite{scells2018querylab}. Each fragment contains at least one MeSH term. This results in a total of 311 unique query fragments (2.68 fragments per query on average). For each query fragment, we corrected any errors (e.g., spelling mistakes, syntactic errors), extracted MeSH terms, keywords, query fragments with MeSH terms, and query fragments without MeSH terms. 
For training the LTR model for each lexical method, we use the pre-split training and test portions from the CLEF datasets. The 2019 topics are also split on systematic review type (intervention and diagnostic test accuracy --- indicated as `intervention' and `dta' respectively in the results), while those for 2017 and 2018 are all diagnostic test accuracy. We use the `quickrank' library~\cite{capannini2016quality} for LTR, instantiated with LambdaMART trained to maximise nDCG. We leave other settings as per default.


For learning the linear function of the BERT suggestion method to decide the cut-off value of the MeSH term ranking list, as described in Section~\ref{sec:BERT_suggester}, we use the training portions of the CLEF TAR datasets. First, we obtain all the \fgms from the CLEF TAR training splits. We count the number of \ftacs and MeSH terms in each \fgm. We then perform linear regression on these numbers to determine a function for each CLEF TAR dataset. We show the linear regression in Figure \ref{fig:learn_J}.

\section{Results}

Results in this section are presented on the test splits of the CLEF TAR datasets (i.e., 2017, 2018, 2019-dta, 2019-intervention). We first analyse the search effectiveness of lexical methods versus our new BERT methods and then analyse the MeSH suggestion effectiveness compared to the MeSH terms originally used.

\subsection{Retrieval Effectiveness}
\label{sec:searcb_effectiveness_nonBERT}

\paragraph{Lexical Methods}
\label{lexi_finding}
The results of the lexical methods presented in Table~\ref{table:search_result} are the same reported in our previous work~\cite{wang2021mesh}. We discuss them briefly here for completeness. Unrefined methods generally contain higher recall than corresponding refined methods, with lower precision. This finding indicates that adding more MeSH terms in the query fragments can cause both more relevant and irrelevant studies to be retrieved. When compared using F1 and F3, the refined methods consistently outperform unrefined methods for each dataset. The U-CUT method achieves the highest effectiveness on CLEF 2017 and 2018, while A-CUT can achieve the highest effectiveness on CLEF 2019 dta; M-CUT for CLEF TAR 2019 intervention dataset in terms of F1 and F3.
%
%
In terms of recall, the unrefined fusion method achieves the highest recall among all lexical suggestion methods. This gain in the recall is likely because unrefined fusion combines all of the MeSH terms suggested by the other three methods (ATM, MetaMap, and UMLS) using `OR'. This suggests that the unrefined fusion method is not beneficial for improving the precision of a Boolean query. However, suppose semi-automatic MeSH term suggestion can be used. Information specialists may be able to use the suggestion and apply their expertise to decide which MeSH terms may be included to achieve higher performance.


\paragraph{BERT Methods}
\label{sec:searcb_effectiveness_BERT}

We first compare the effectiveness of the BERT methods with the original Boolean query. The result shows that under all evaluation measures (Precision, F1, F3 and Recall), BERT methods can outperform the original query on CLEF TAR 2017, 2018 and 2019-intervention, while effectiveness is generally worse for CLEF TAR 2019-dta. Note that CLEF TAR 2019-dta only contains eight unique topics; the lower effectiveness is likely due to a handful of topics.

Next, we compare the effectiveness of BERT suggestions against lexical suggestions. When comparing with un-refined lexical methods, the effectiveness of BERT suggestions is comparable in terms of F1 and F3, showing substantial gains across all datasets. However, compared with refined lexical methods, BERT suggestions generally obtain comparable results to refined lexical suggestions, except in CLEF TAR 2019-dta, in which refined lexical suggestion methods achieve higher effectiveness.
In terms of recall, BERT suggestions obtain slightly higher recall to un-refined lexical suggestions, but substantially higher recall than refined lexical suggestions.

As mentioned in Section \ref{lexi_finding}, unrefined lexical methods are effective to achieve higher recall while refined lexical methods are effective to achieve higher Precision, F1 and F3. We find that the MeSH terms suggested by BERT can obtain similar recall effectiveness to un-refined lexical methods while F1 and F3 can be comparable to refined lexical methods. Therefore, compare with lexical methods, BERT methods may be preferred to suggest more effective MeSH Terms.




\subsection{Impact of BERT Ranking Representations}
We compare different ranking representations of BERT, including \atb, \seb and \frb. We use the same cut-off strategy to compare these three representations fairly. We find that the precision, F1 and F3 values of \frb are the highest among the three methods, while recall of \frb is the lowest. However, only one MeSH term is suggested for each \fgm when \frb is used. This trade-off of precision and recall also suggests the same finding we described for lexical methods, where adding more MeSH terms can cause more studies to be retrieved.
Between \seb and \atb, \seb is able to obtain higher precision while recall is lower than \atb. When comparing using F1 or F3, \seb always achieves higher effectiveness. %
Therefore, the use of \seb is preferred over \atb.


\subsection{Impact of Cut-off Strategy}
When comparing different cut-off strategies for BERT suggestions\footnote{Only \frb is considered as SA, SO, LN are only applicable to \frb.}, we find that FO can consistently achieve the highest Precision, F1 and F3 compared to other cut-off methods. On the other hand, the recall value of using FO is the lowest among all other methods, indicating that the trade-off of precision and recall is again caused by the number of MeSH terms added to the query. For the other three cut-off strategies, including SA, SO, and LN, we find that SO and LN consistently outperform SA, suggesting that information specialists have an intuition for how many MeSH terms to add to a query.

\subsection{Are Suggested MeSH Terms the Same as those in the Original Queries?}

\begin{table*}
	\centering
	\small
	\begin{tabular}{p{2pt}l|cc|cc|cc|cc}
    \\ \toprule
		\multicolumn{2}{c|}{Dataset}&\multicolumn{2}{c|}{2017}&\multicolumn{2}{c|}{2018}&\multicolumn{2}{c|}{2019-dta}&\multicolumn{2}{c}{2019-intervention}\\ \midrule
		
		\multicolumn{2}{c|}{Method}&\multicolumn{1}{c}{Jaccard}&\multicolumn{1}{c|}{Num}&\multicolumn{1}{c}{Jaccard}&\multicolumn{1}{c|}{Num}&\multicolumn{1}{c}{Jaccard}&\multicolumn{1}{c|}{Num}&\multicolumn{1}{c}{Jaccard}&\multicolumn{1}{c}{Num}\\ \midrule

\multirow{8}{*}{\rotatebox{90}{Lexical Method}}&ATM&0.0999&5.5373&0.2368&6.0139&0.2117&5.1500&0.2356&4.8868\\
&ATM-CUT&0.1995$^{*}$&2.4179&0.1938&2.3056&0.2004&2.0500&0.2109&1.3019\\
&MetaMap&0.2654$^{*}$&4.6866&0.2218&4.0417&0.2163&4.8000&0.2069&4.5094\\
&MetaMap-CUT&0.2374$^{*}$&2.3134&0.1964&1.9028&0.2241&2.3500&0.1981&1.7736\\
&UMLS&0.2243$^{*}$&8.9254&0.2235&7.9722&0.1905&7.7000&0.2405&7.5660\\
&UMLS-CUT&0.2751$^{*}$&1.8955&0.2424&1.8611&0.1986&2.2000&0.2050&1.7547\\
&Fusion&0.2165$^{*}$&11.4776&0.2160&10.9444&0.1735&10.5000&0.2212&9.7358\\
&Fusion-CUT&0.2761$^{*}$&2.7761&0.2742&3.3194&0.2508&3.1000&0.2909&2.4340\\\midrule
\multirow{6}{*}{\rotatebox{90}{BERT Method}}&Atomic-BERT-FO&0.2532$^{*}$&12.7313&0.3105&12.2639&0.1573&11.8500&0.2252&13.6226\\
&Semantic-BERT-FO&0.2370$^{*}$&11.0746&0.2963&10.6944&0.1654&10.7500&0.2219&11.5283\\
&Fragment-BERT-FO&0.3455$^{*}$&1.0000&0.3812$^{*}$&1.0000&0.1681&1.0000&0.2235&1.0000\\
&Fragment-BERT-SA&0.2233$^{*}$&\textbf{16.6269}&0.2639&\textbf{16.4861}&0.1790&\textbf{15.5000}&0.2531&\textbf{17.2264}\\
&Fragment-BERT-SO&\textbf{0.3921$^{*}$}&4.1343&\textbf{0.4634$^{*}$}&4.8333&0.2574&4.4000&\textbf{0.3301}&2.7547\\
&Fragment-BERT-LN&0.2780$^{*}$&5.2687&0.2689&3.7778&\textbf{0.2667}&3.8500&0.2415&3.8491\\

		\bottomrule
	\end{tabular}
	\caption{Jaccard index(Jaccard) values quantifying the overlap between the MeSH terms suggsted by the investigated methods and those in the original query, along with the average number (Num) of MeSH term suggested by each method. In the original queries, there were on average 4.1343 MeSH terms for 2017, 4.8333 for 2018, 4.4000 for 2019-dta, and 2.7547 for 2019-intervention. Lexical methods: \textit{CUT} indicates cut-off ranks.  BERT methods: \textit{FO}, \textit{SA}, \textit{SO}, \textit{LN} indicate different cut-off strategies. Two-tailed statistical significance (t-test, $p<0.05$) with Bonferroni correction between ATM and the other methods is indicated by $*$.}	
	\label{table:suggestion_result}
	
\end{table*}


Next, we study the overlap between the MeSH term suggested by the considered methods and those included in the original query; this is reported in Table~\ref{table:suggestion_result} and is measured with the Jaccard index. One immediate observation is that the overlap of all based methods is considerably higher than that of lexical methods. This observation is based on that the highest value of the Jaccard index in each dataset always appear in the BERT suggestion method. Moreover, when applying the SO cut-off strategy to \frb, the highest overlap is always obtained, which indicates that BERT suggestion methods also agree on the Terms chosen by systematic reviewers.  .

\begin{figure}[!thb]
	
	\begin{minipage}{0.245\textwidth}
		\centering
		\includegraphics[width=1\linewidth]{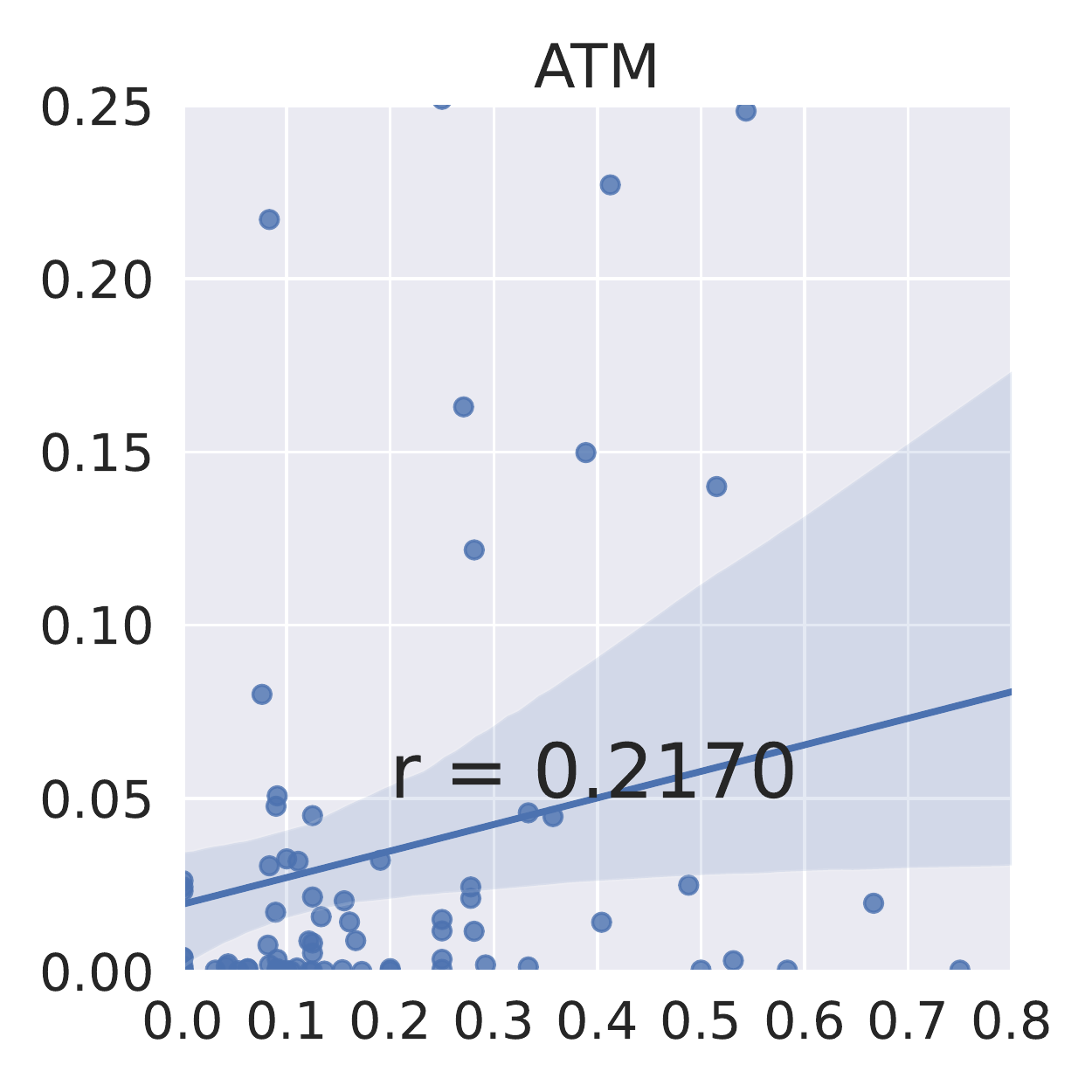}
	\end{minipage}\hfill
	\begin{minipage}{0.245\textwidth}
		\centering
		\includegraphics[width=1\linewidth]{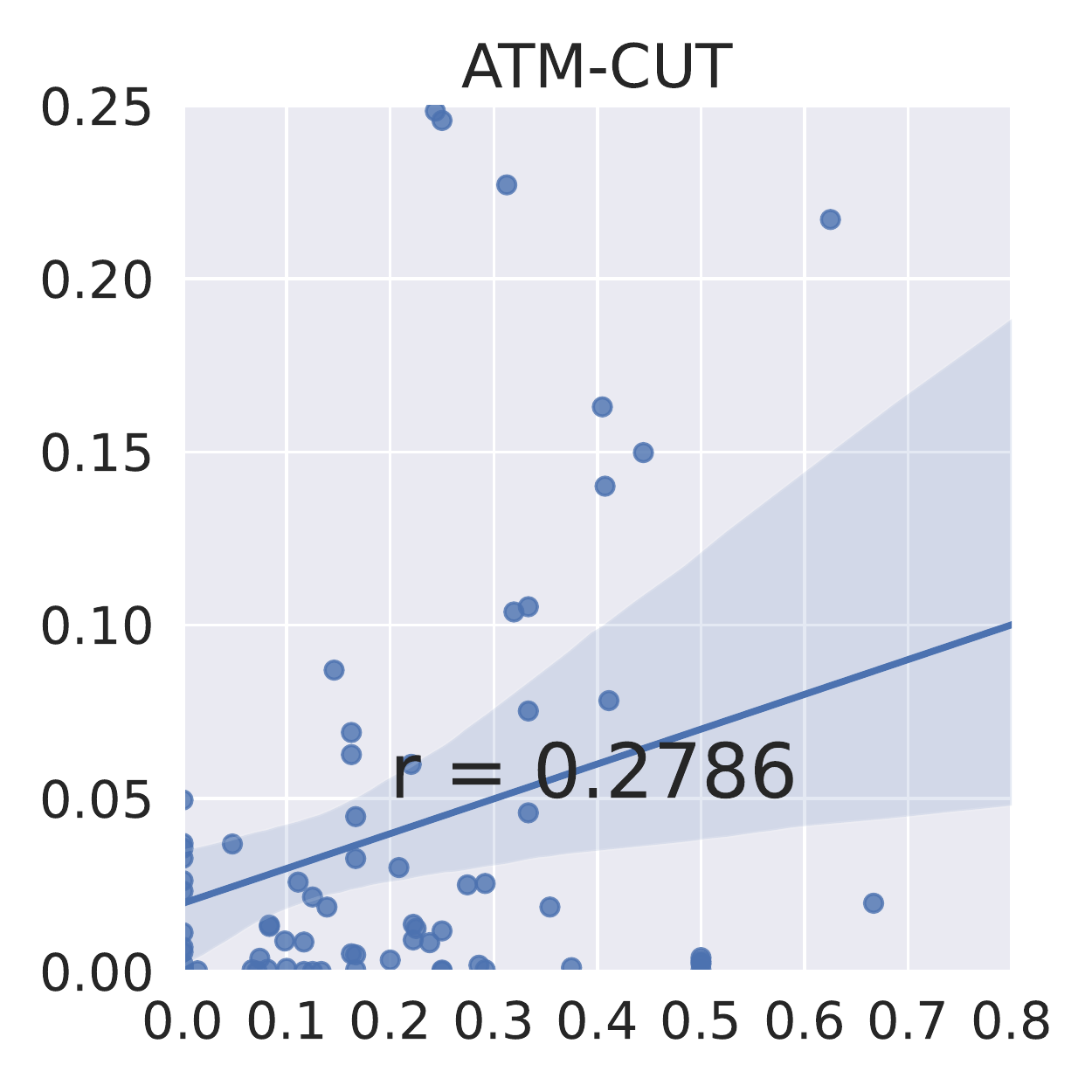}
	\end{minipage}
	\begin{minipage}{0.245\textwidth}
		\centering
		\includegraphics[width=1\linewidth]{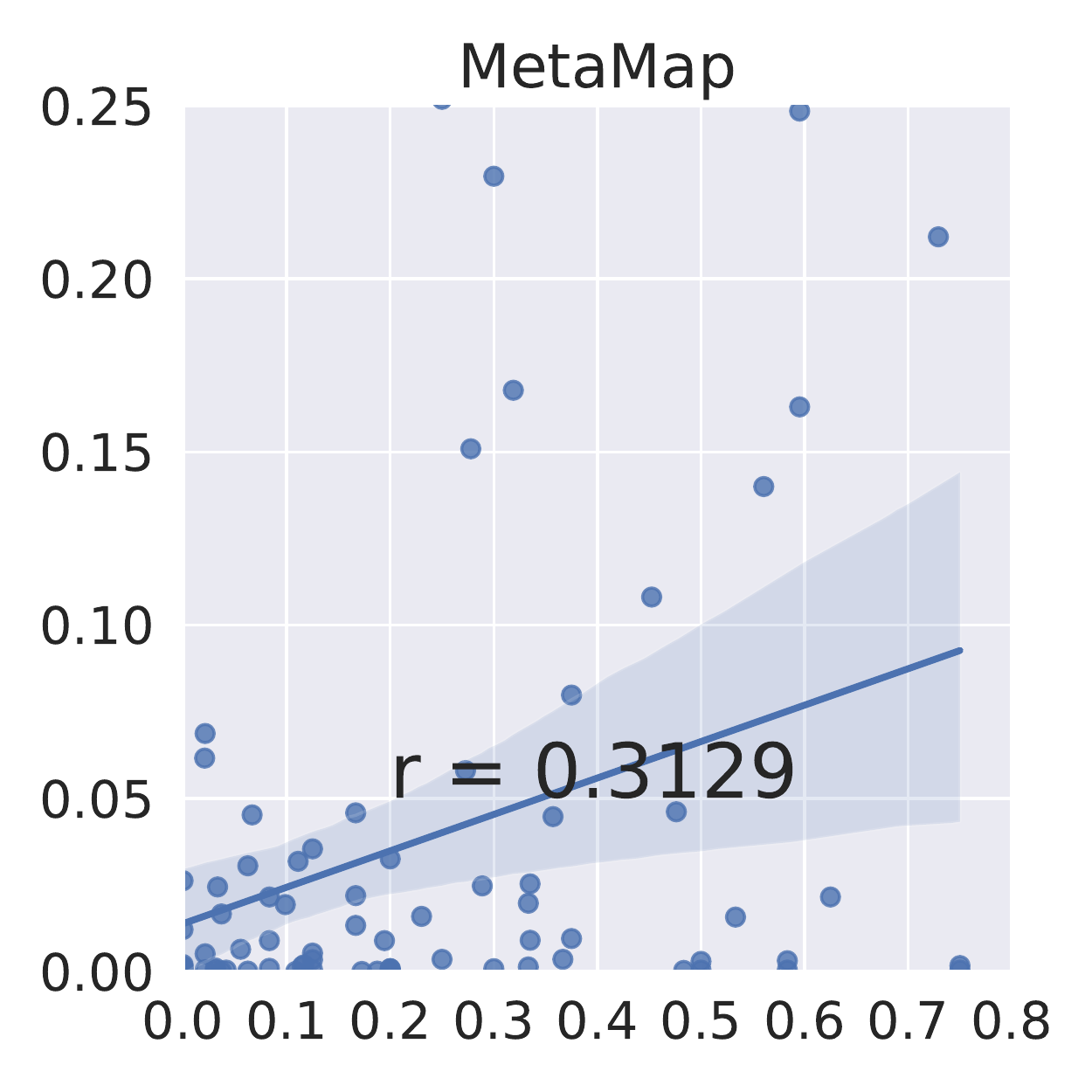}
	\end{minipage}
	\begin{minipage}{0.245\textwidth}
		\centering
		\includegraphics[width=1\linewidth]{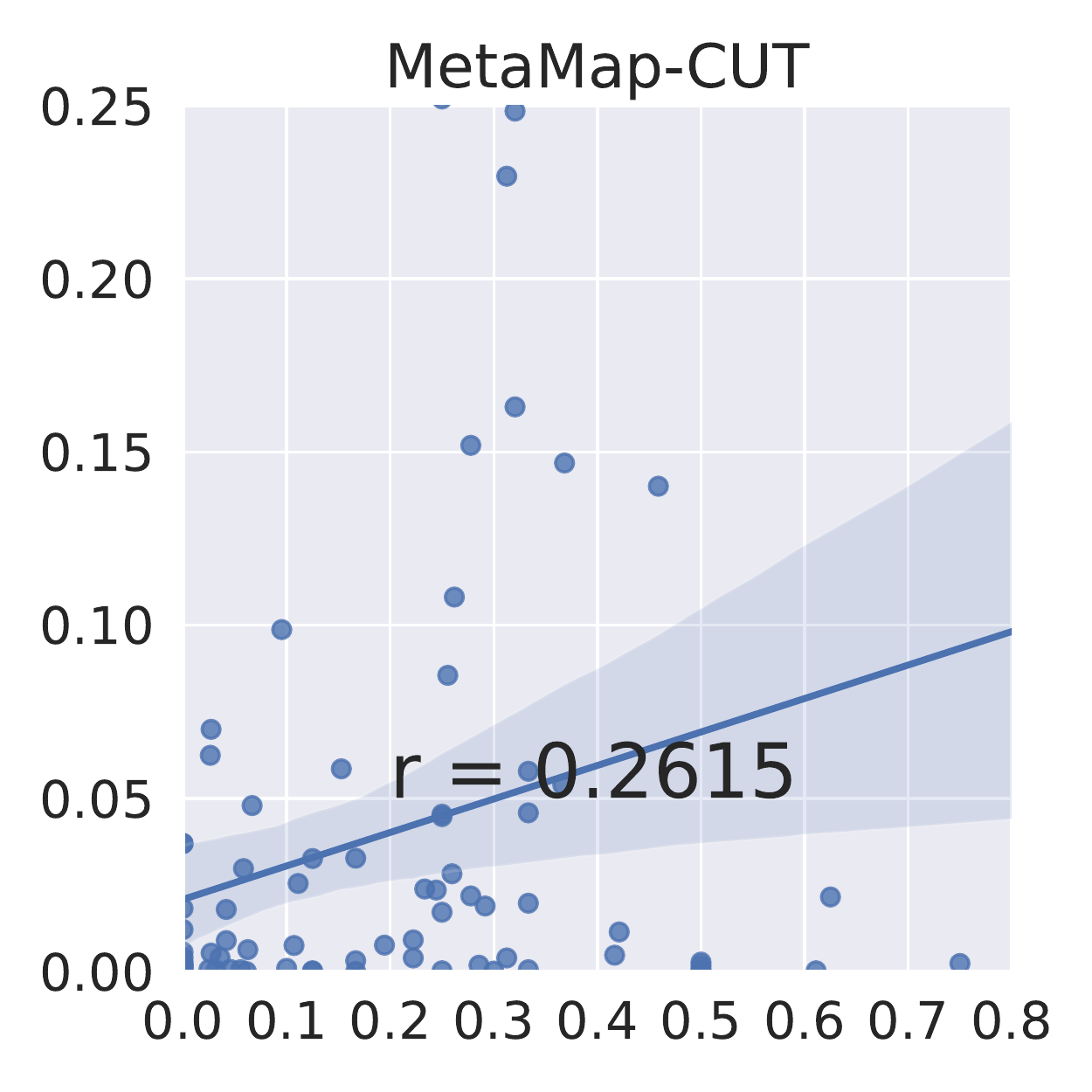}
	\end{minipage}
	\begin{minipage}{0.245\textwidth}
		\centering
		\includegraphics[width=1\linewidth]{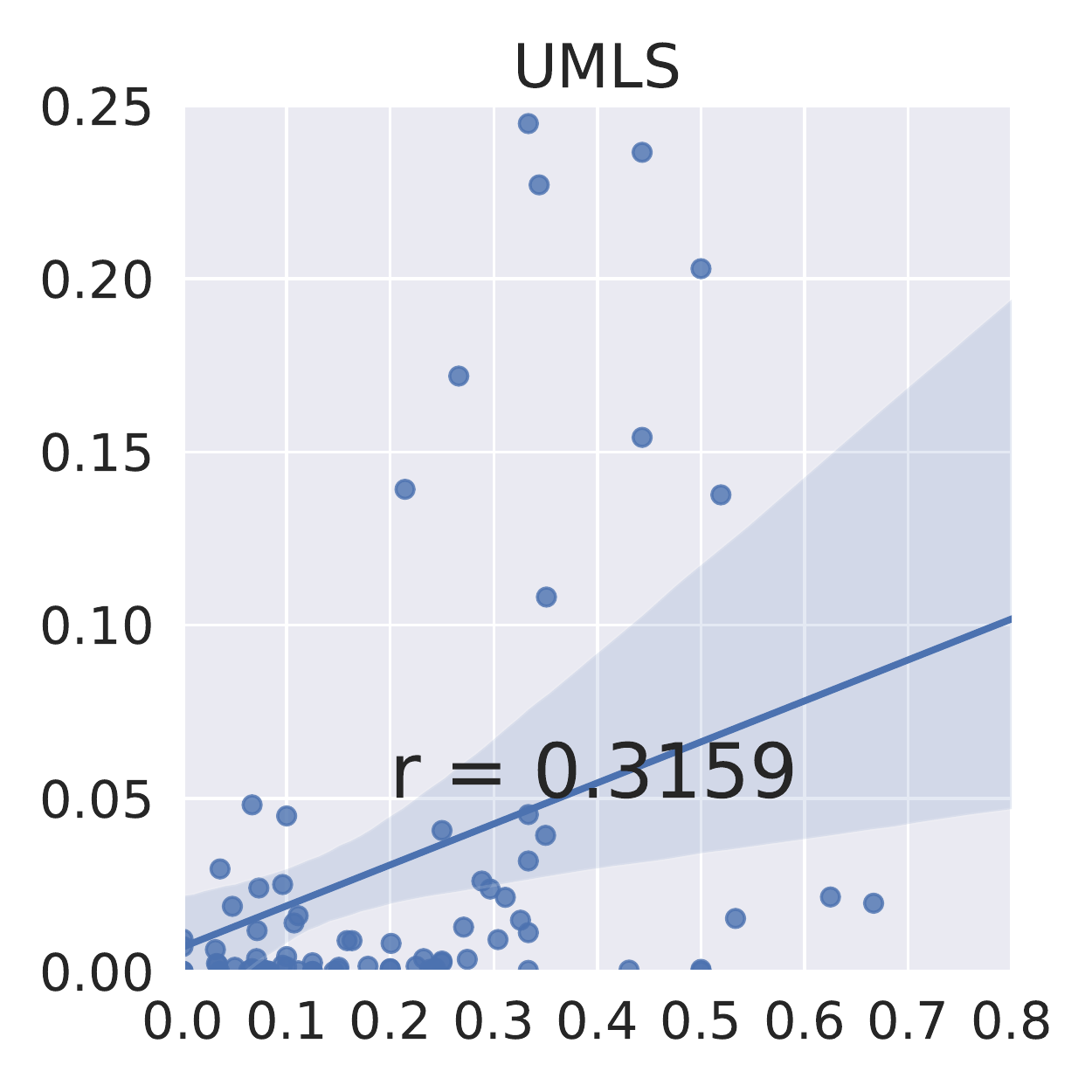}
	\end{minipage}
	\begin{minipage}{0.245\textwidth}
		\centering
		\includegraphics[width=1\linewidth]{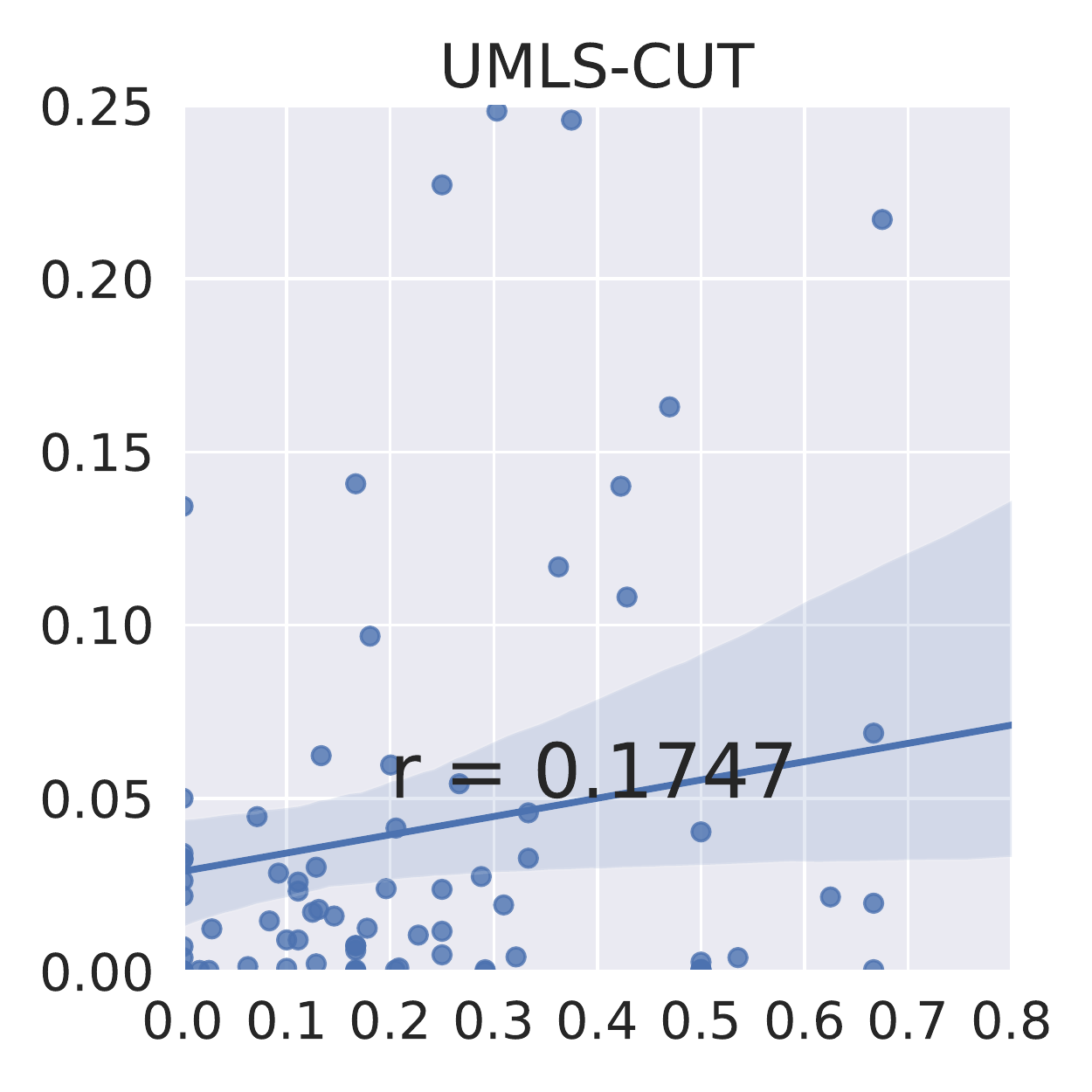}
	\end{minipage}
	\begin{minipage}{0.245\textwidth}
		\centering
		\includegraphics[width=1\linewidth]{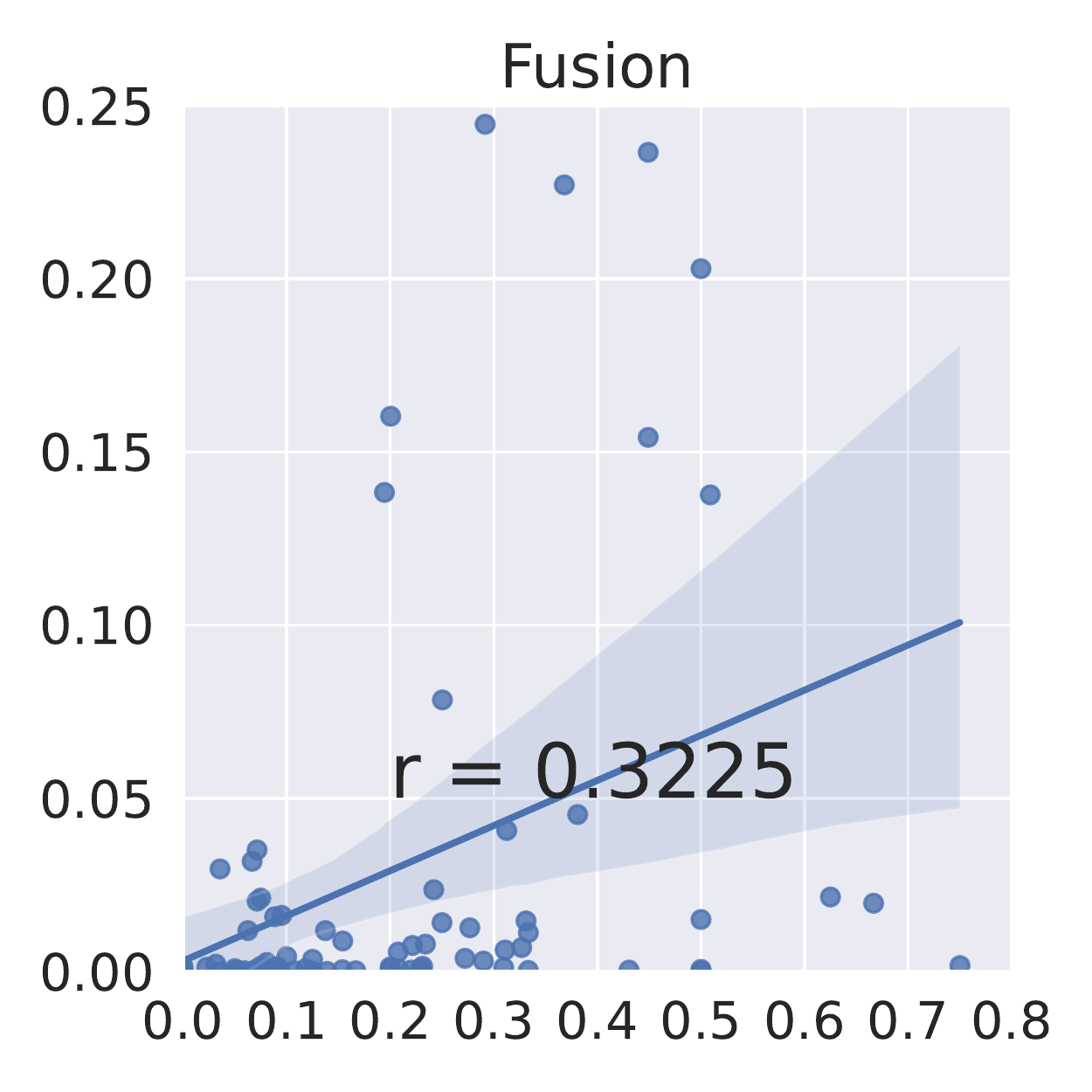}
	\end{minipage}
	\begin{minipage}{0.245\textwidth}
		\centering
		\includegraphics[width=1\linewidth]{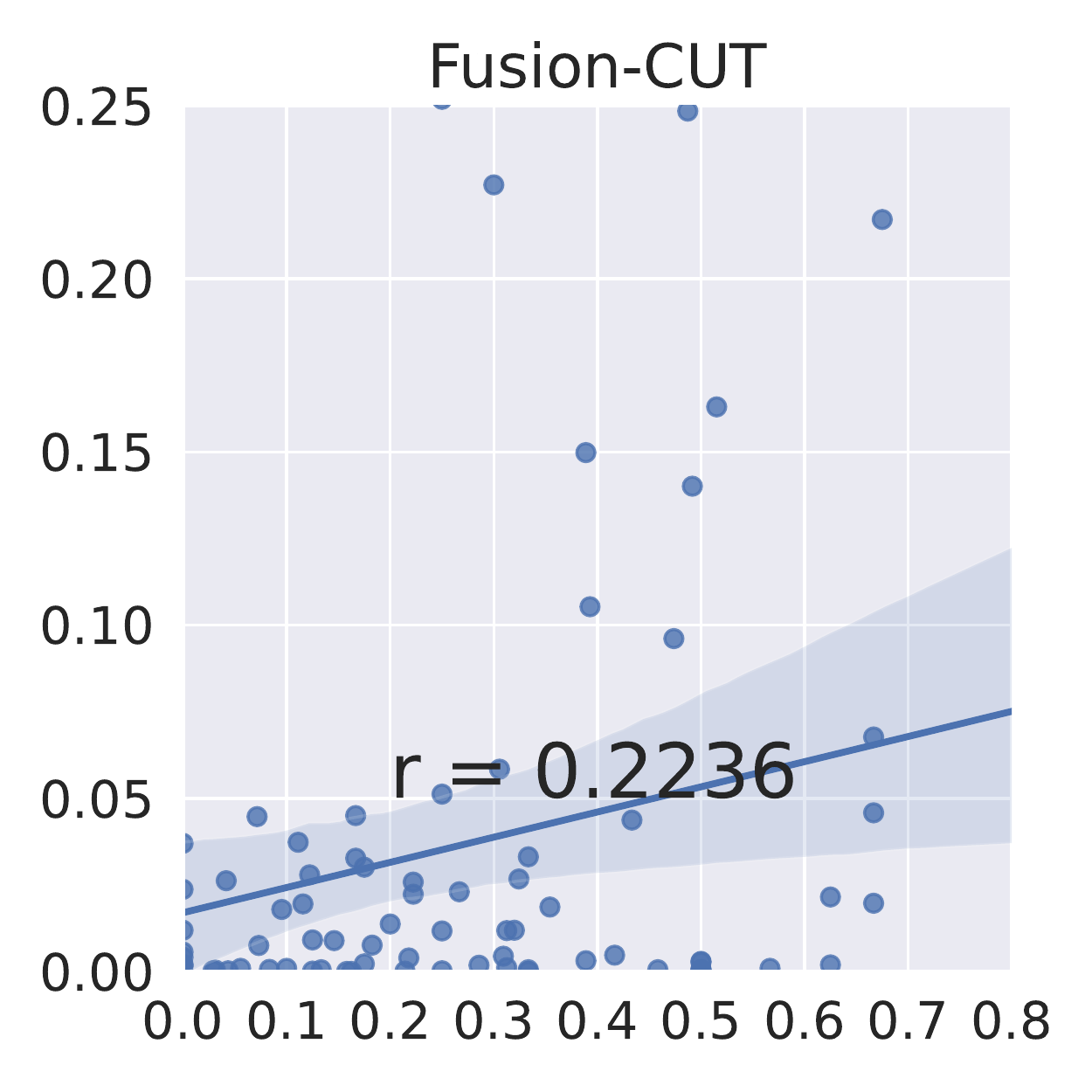}
	\end{minipage}
	\begin{minipage}{0.245\textwidth}
		\centering
		\includegraphics[width=1\linewidth]{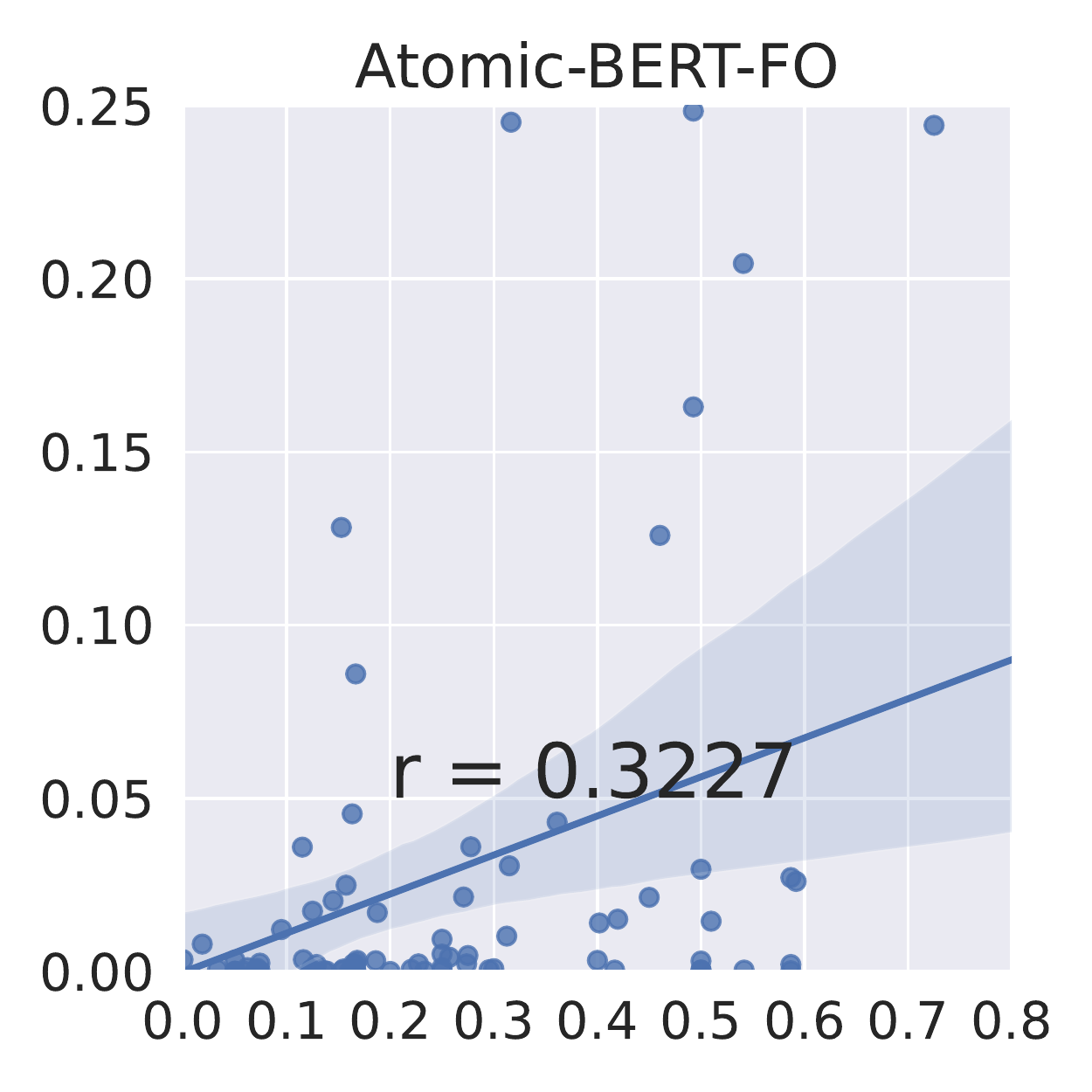}
	\end{minipage}\hfill
	\begin{minipage}{0.245\textwidth}
		\centering
		\includegraphics[width=1\linewidth]{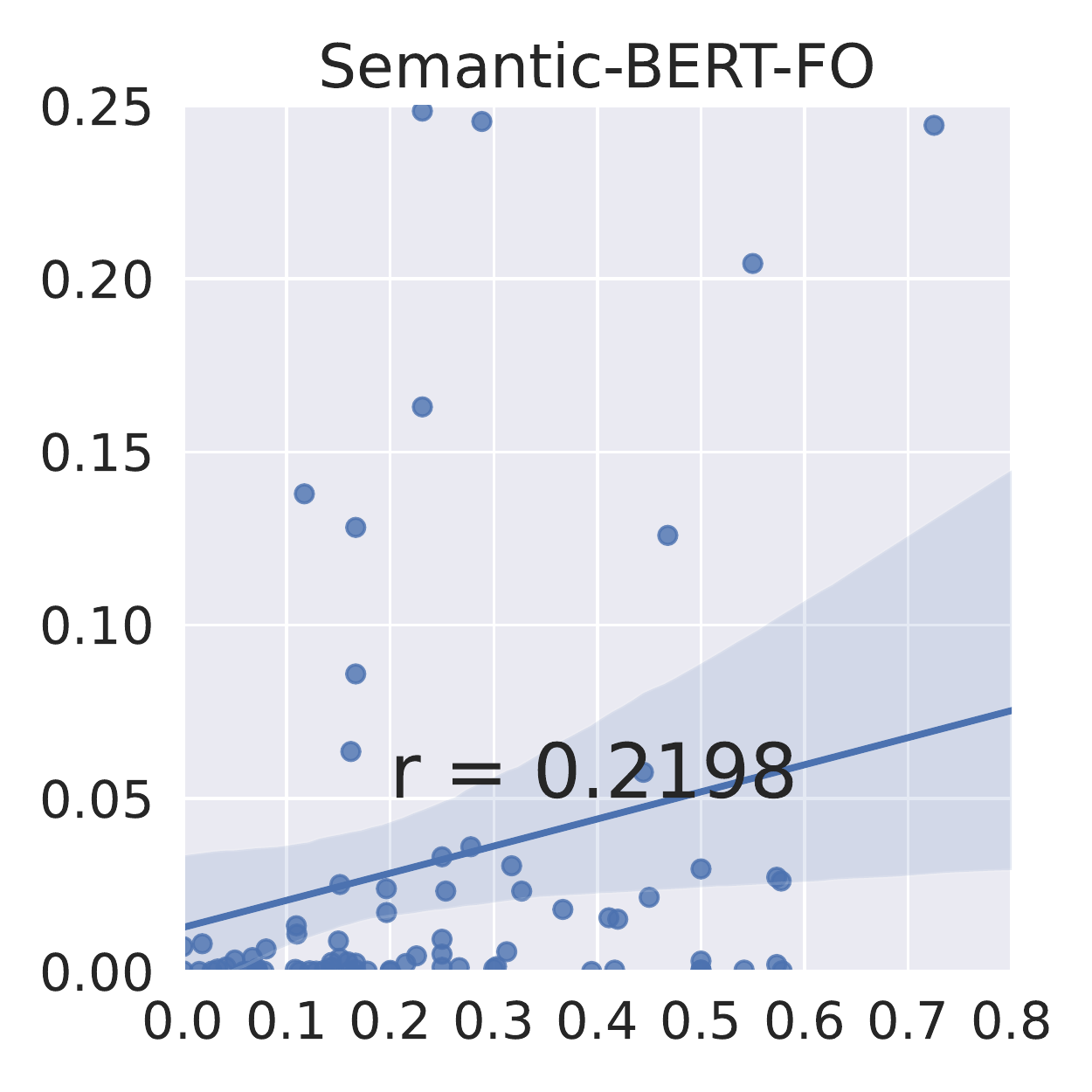}
	\end{minipage}\hfill
	\begin{minipage}{0.245\textwidth}
		\centering
		\includegraphics[width=1\linewidth]{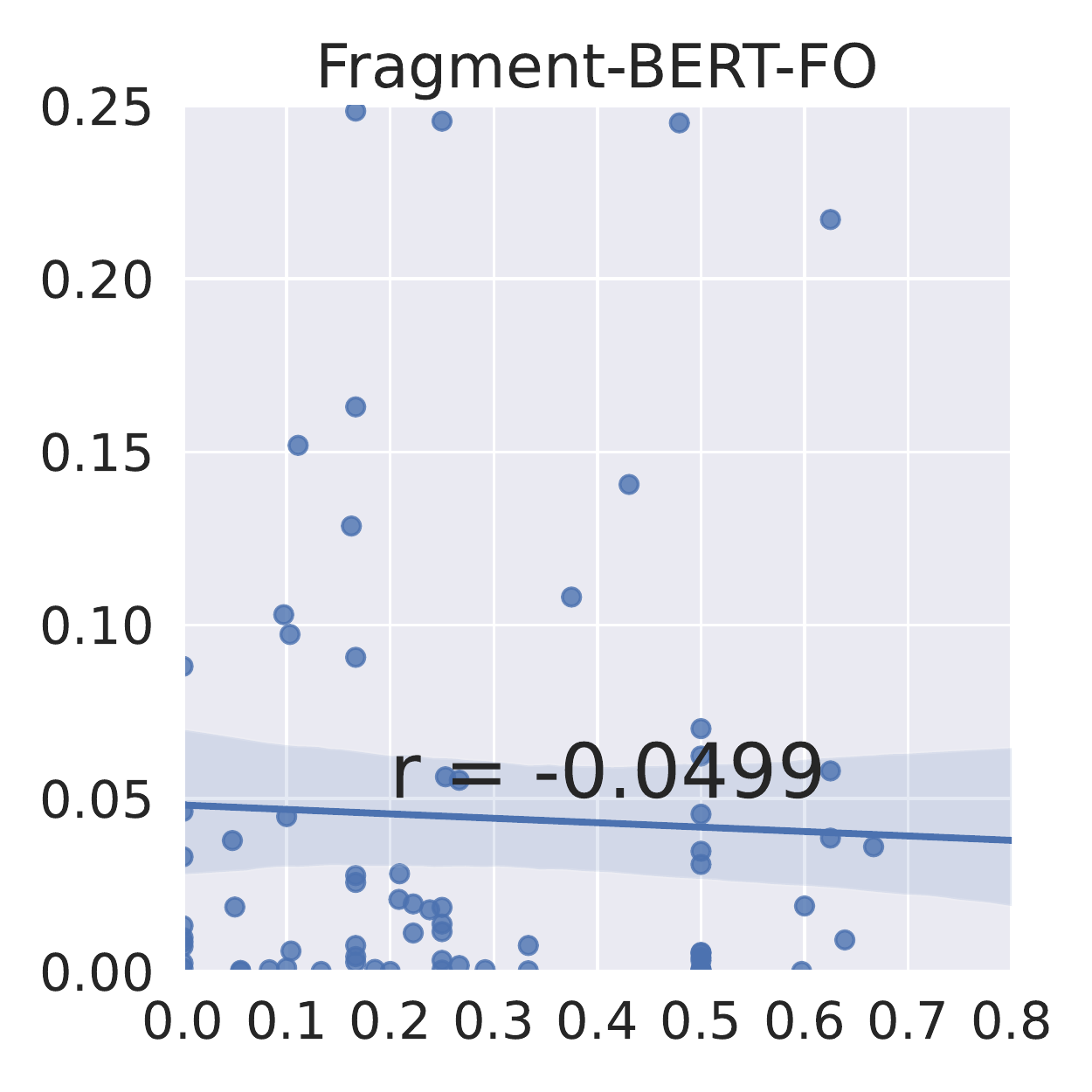}
	\end{minipage}\hfill
	\begin{minipage}{0.245\textwidth}
		\centering
		\includegraphics[width=1\linewidth]{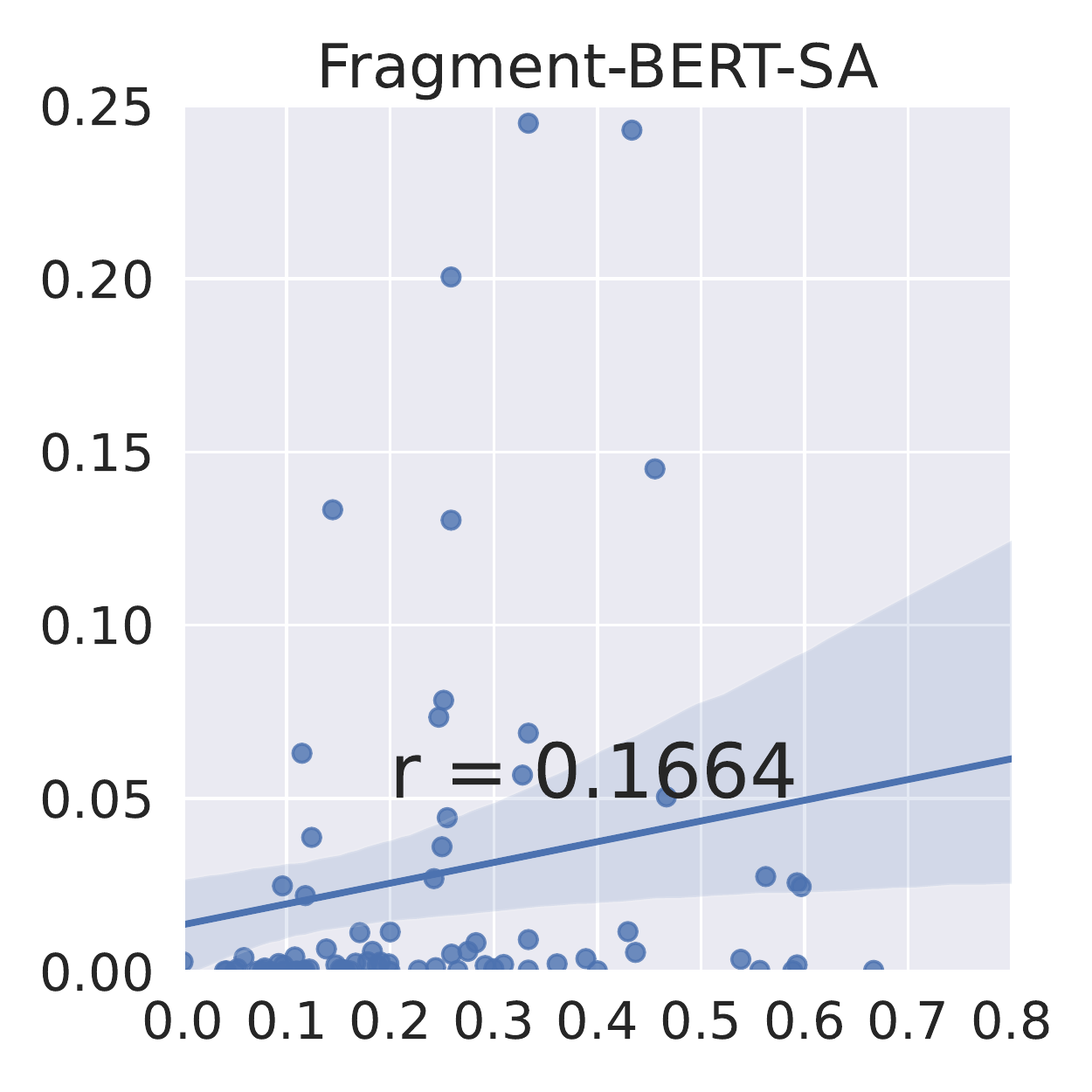}
	\end{minipage}
	\begin{minipage}{0.245\textwidth}
		\centering
		\includegraphics[width=1\linewidth]{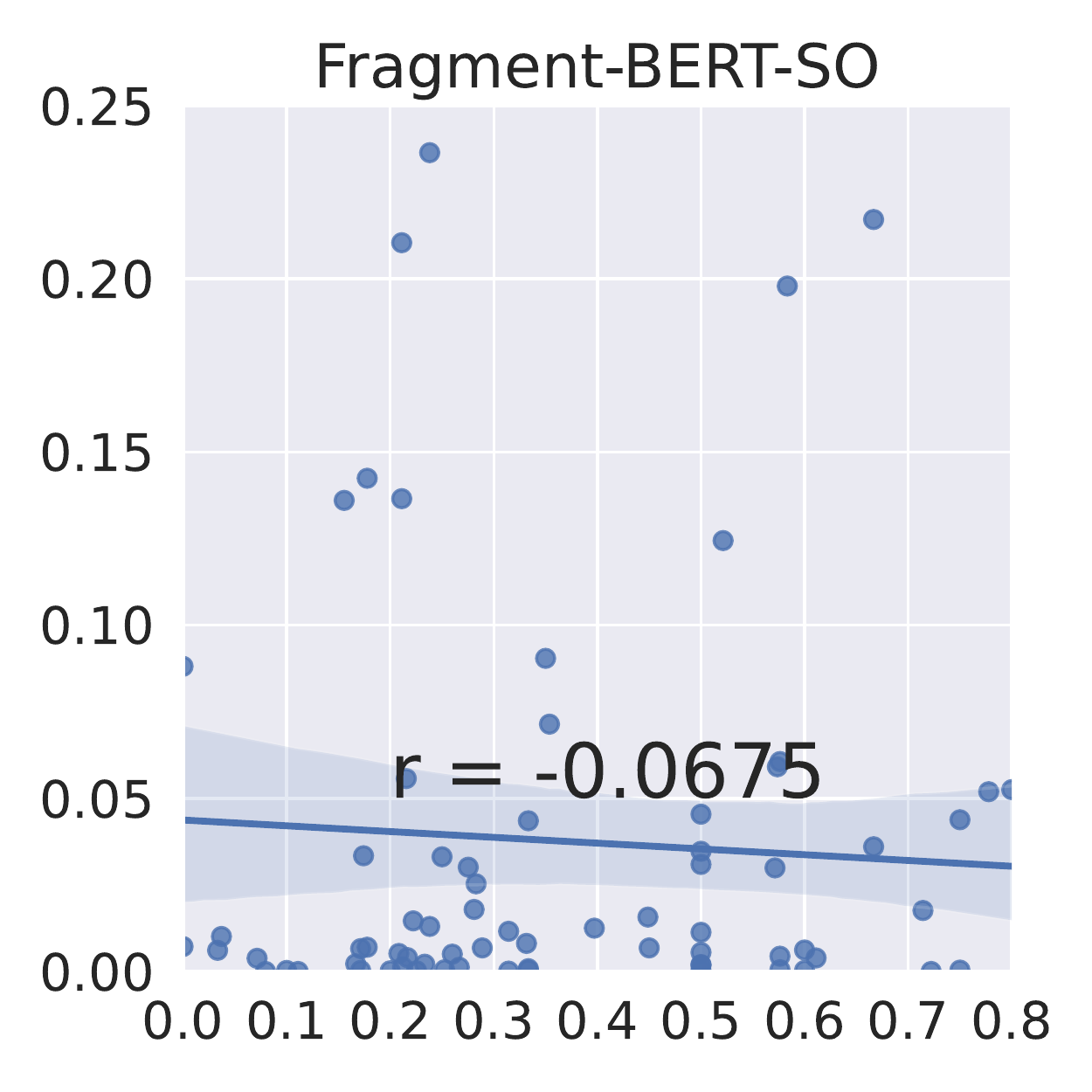}
	\end{minipage}
	\begin{minipage}{0.001\textwidth}
		\centering
	\end{minipage}
	\begin{minipage}{0.245\textwidth}
		\centering
		\includegraphics[width=1\linewidth]{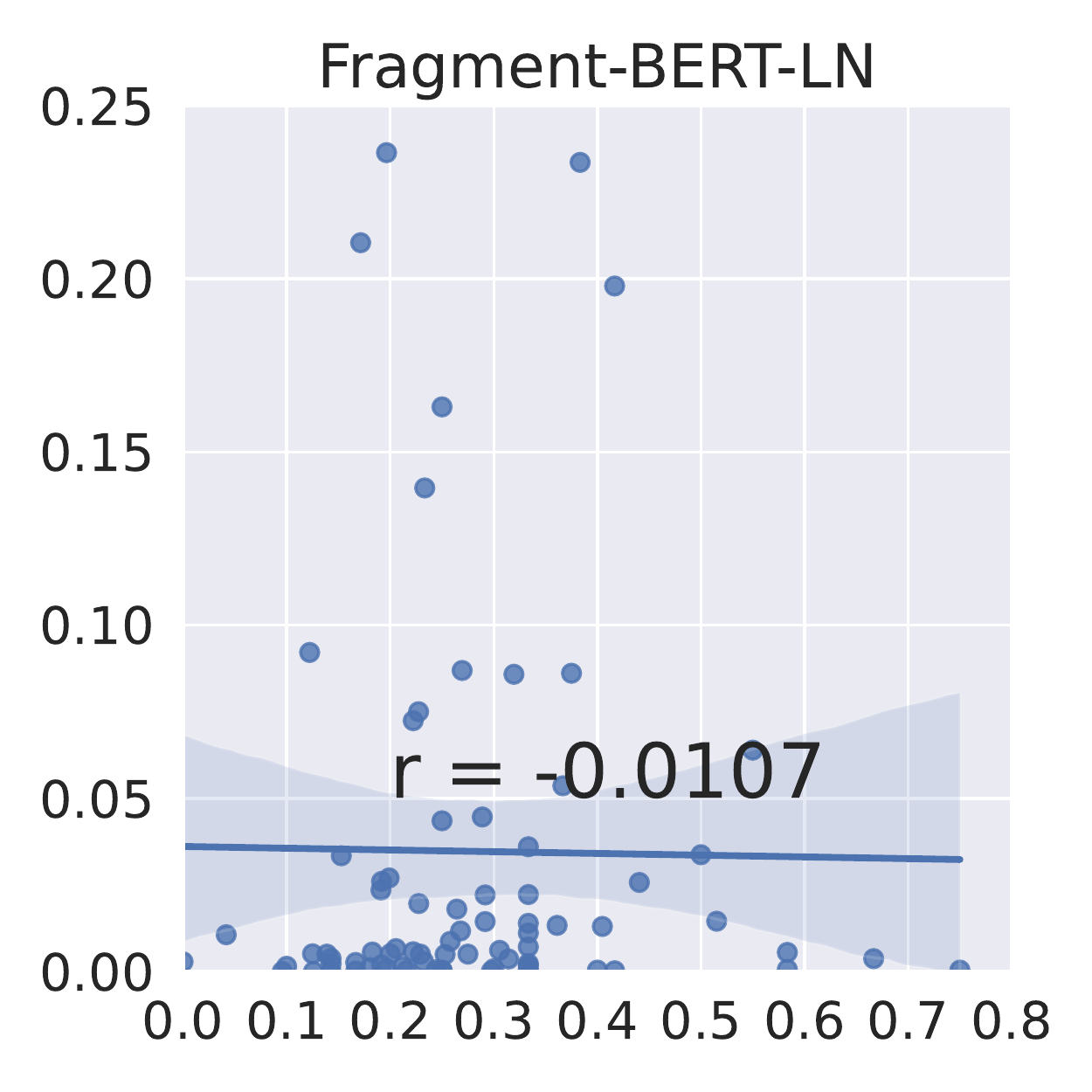}
	\end{minipage}
	
	\caption{Correlation graph of search effectiveness versus the overlap of MeSH terms. The x-axis reports the Jaccard index for overlap between suggested MeSH terms and MeSH terms included in the original query, and the y-axis reports the F1 values for search effectiveness for each topic.}
	\label{fig:correlation}
	
\end{figure}

The previous results reported in Table~\ref{table:search_result} highlighted that, in general, BERT methods were better than lexical methods in suggesting effective search terms; and these were more effective than those in the original queries, although differences were not statistically significant. These results, in conjunction with the findings in Table~\ref{table:suggestion_result}, indicate that the BERT methods identify very similar MeSH terms that present in the original queries -- and the MeSH terms identified by BERT methods are more effective than those provided by other methods.

Intuitively, we further analyse whether search effectiveness and the suggestion of MeSH terms that are included in the original query correlate, meaning that Mesh Terms used in the original Boolean query may be of very high quality and should be used as gold standard. The Jaccard index measure is used once more to represent the similarity between suggested MeSH terms and those present in the original query, and F1 is used to represent the search effectiveness of the associated query (with suggested MeSH terms included).
The results of this correlation analysis are reported in Fig~\ref{fig:correlation}. We find that, while for all lexical methods search effectiveness is weakly correlated with the overlap of MeSH terms, this is not the case for BERT methods. This indicates that \mts from the original query may not be the best \mts to suggest. In fact, it often is that MeSH terms that are suggested but not included in the original query provide higher search effectiveness than the original MeSH terms themselves.



\begin{figure}[!thb]
	
	\begin{minipage}{\textwidth}
		\centering
		\includegraphics[width=1\linewidth]{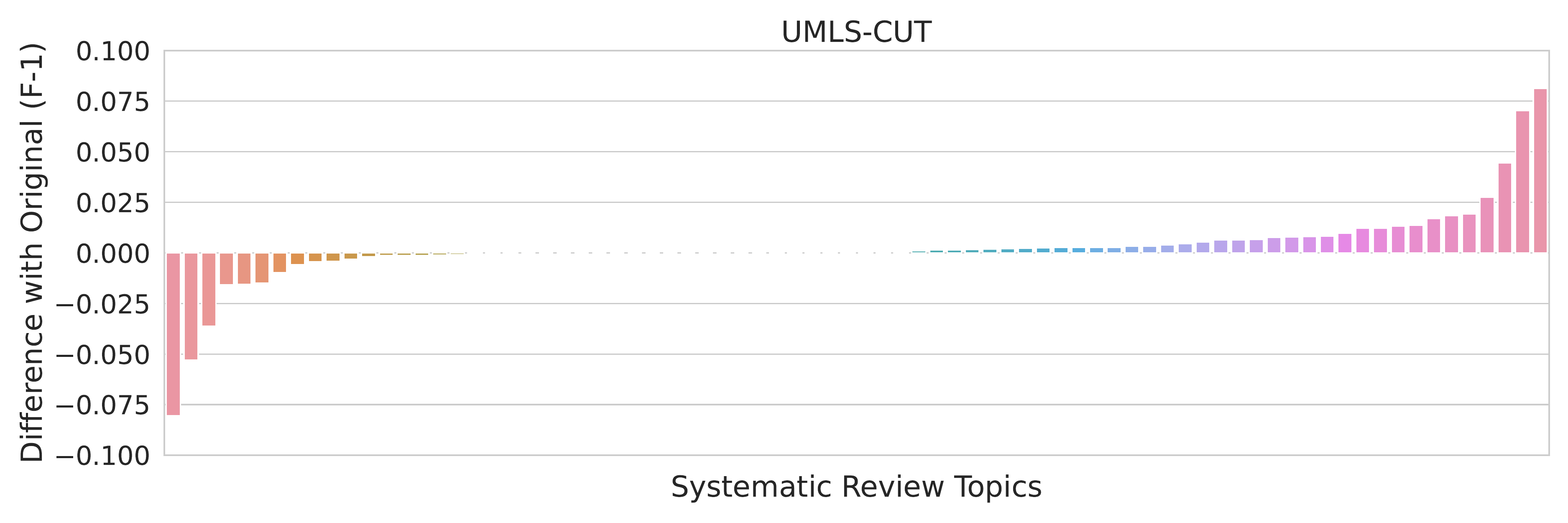}
	\end{minipage}
	
	\begin{minipage}{\textwidth}
		\centering
		\includegraphics[width=1\linewidth]{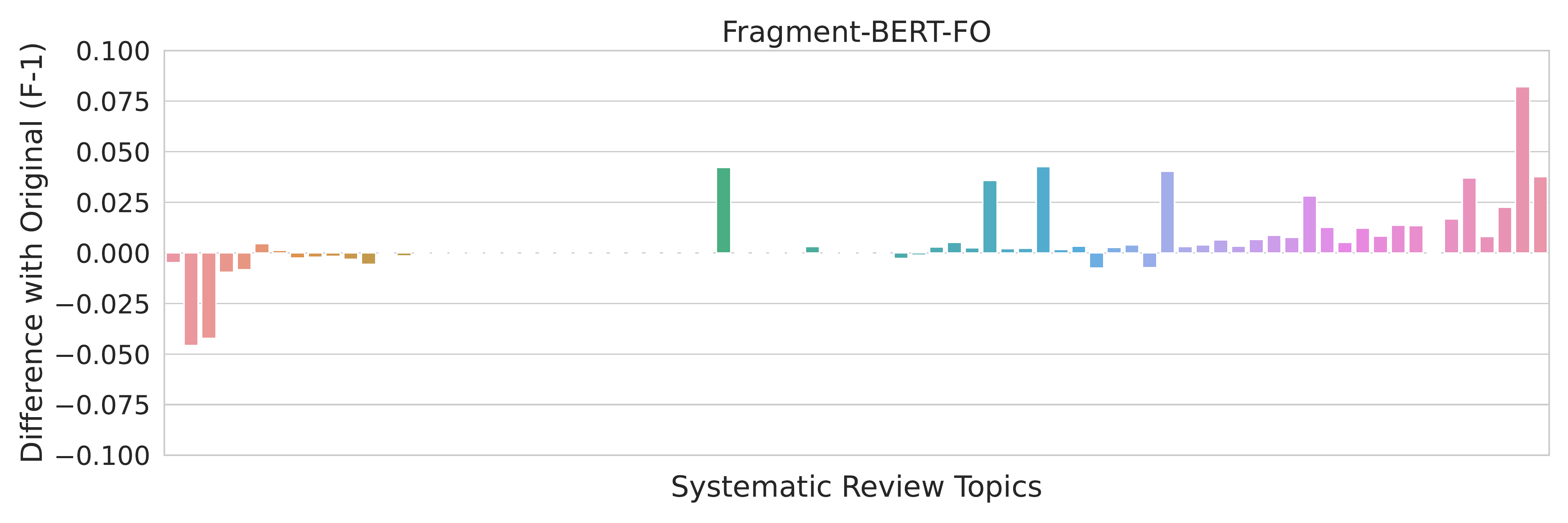}
	\end{minipage}
	\caption{Plot showing systematic review topics versus original query effectiveness; each bar represents a topic. The y-axis represents the effectiveness difference between the query with the suggested MeSH terms and the original query. Effectiveness is measured using F1.}
	\label{fig:search_stability}
\end{figure}

\subsection{Search Stability}
We next analyse the search effectiveness stability of different MeSH term suggestion methods on a topic-by-topic basis. With search effectiveness stability we refer to the amount of variance across topics of the measured search effectiveness obtained when using queries with MeSH terms suggested by a specific MeSH term suggestion method. The larger the effectiveness, the lower the stability.
We only analyse the best-performing lexical (U-CUT) and BERT (F-B-FO) methods.

Figure~\ref{fig:search_stability}, which combines the topics of all of the CLEF TAR datasets test splits, shows that for most of the topics, both kinds of MeSH suggestion methods can outperform or match the effectiveness of the original queries.
We also find that our MeSH term suggestion methods sometimes obtain lower effectiveness. It is unclear if these are difficult topics to suggest MeSH terms for, or if there are mistakes in these queries that cause the poor effectiveness (e.g., spelling mistakes in the \ftacs that were not detected at the time of data cleaning). 

\subsection{Case Study}

Given the findings above, we next seek to investigate the reasons for highly effective or ineffective results. 
We choose topic CD009642 and topic CD004414 from the CLEF TAR 2019-intervention dataset as they are representative topics where suggestion methods outperform the effectiveness (CD009642) or struggle to match the effectiveness (CD004414) of the original query. 
Query fragments corresponding to these topics for all the suggestion methods are shown in Tables~\ref{table:casestudy_better} and~\ref{table:casestudy_worse}. 
We also show their search effectiveness in Table \ref{table:case_resulta}.


Firstly, we find that both the suggested MeSH terms and the search effectiveness are similar for all lexical methods. One exception is UMLS in topic CD004414, which suggests more MeSH terms than the other two methods, causing a drop in effectiveness.

On the other hand, MeSH terms suggested by BERT methods appear to differ greatly from lexical methods. 
BERT methods have captured both lexically similar MeSH terms and terms semantically related to the input \ftacs. One example is shown in the first fragment of topic CD009642. While all lexical methods suggest \textit{Lidocaine} which is lexically equal to \textit{lidocain}, BERT methods suggest similar drugs such as \textit{Procaine}. Another example in topic CD004414 shows that BERT methods can use this semantic matching ability to suggest MeSH terms indicating the method of intervention, shown in suggesting \textit{Patch Tests}. Therefore, BERT methods suggest MeSH terms that are not bound to the lexical semantics of a \ftac. Another advantage of BERT methods is that they guarantee that at least one MeSH term will be suggested. For lexical methods, suggestions are based on pre-existing rule-based knowledge; thus, when \ftacs can not be matched, no MeSH terms can be suggested (e.g., ATM does not suggest any MeSH term in Fragment 1 of CD009642). Overall, we believe that the semantic matching of BERT may sometimes be detrimental to MeSH suggestion. We leave the investigation into how to prevent BERT from suggesting MeSH terms that are not relevant to the information need of a query fragment (e.g., suggesting a MeSH term that is the intervention of an outcome for a query fragment) for future work.


\begin{table}[t!]
	\centering
	\normalsize
	
	\begin{tabular}{l|p{36pt}|p{36pt}|p{36pt}|p{36pt}|p{36pt}|p{36pt}|p{36pt}|p{36pt}}
		\multicolumn{9}{c}{}\\ \toprule
		Topic ID&\multicolumn{4}{c|}{CD009642}&\multicolumn{4}{c}{CD004414}\\\midrule
		Method&P&F1&F3&R&P&F1&F3&R\\ \midrule
		ORIGINAL& 0.0088& 0.0175& 0.0344& 1.0000& 0.0013& 0.0026& 0.0052& 0.6875\\\midrule
		ATM& 0.0109& 0.0215& 0.0421& 0.9194& 0.0018& 0.0035& 0.0070& 0.3125\\
		ATM-CUT& 0.0109& 0.0215& 0.0421& 0.9194& 0.0020& 0.0040& 0.0078& 0.3125\\
		MetaMap& 0.0109& 0.0215& 0.0421& 0.9194& 0.0018& 0.0035& 0.0070& 0.3125\\
		MetaMap-CUT& 0.0109& 0.0215& 0.0421& 0.9194& 0.0014& 0.0027& 0.0054& 0.3125\\
		UMLS& 0.0109& 0.0215& 0.0421& 0.9194& 0.0013& 0.0025& 0.0050& 0.3125\\
		UMLS-CUT& 0.0109& 0.0215& 0.0421& 0.9194& 0.0020& 0.0040& 0.0078& 0.3125\\
		Fusion& 0.0109& 0.0215& 0.0421& 0.9194& 0.0018& 0.0035& 0.0069& 0.3125\\
		Fusion-CUT& 0.0109& 0.0215& 0.0421& 0.9194& 0.0014& 0.0027& 0.0054& 0.3125\\\midrule
		Atomic-BERT-FO& 0.0108& 0.0214& 0.0418& 0.9194& 0.0012& 0.0024& 0.0048& 0.3125\\
		Semantic-BERT-FO& 0.0108& 0.0214& 0.0418& 0.9194& 0.0012& 0.0024& 0.0048& 0.3125\\
		Fragment-BERT-SA& 0.0259& 0.0504& 0.0955& 0.9194& 0.0012& 0.0024& 0.0048& 0.3125\\
		Fragment-BERT-SO& 0.0270& 0.0525& 0.0993& 0.9194& 0.0028& 0.0055& 0.0109& 0.3125\\
		Fragment-BERT-LN& 0.0276& 0.0536& 0.1013& 0.9194& 0.0013& 0.0026& 0.0052& 0.3125\\
		\bottomrule
	\end{tabular}
	\caption{Search effectiveness of the Boolean queries with the suggested MeSH terms evaluated by precision (P), F1, F3 and recall (R). For Lexical methods: \textit{CUT} indicates cut-off ranks. BERT methods: \textit{FO}, \textit{SA}, \textit{SO}, \textit{LN} indicate different cut-off strategies.}	
	\label{table:case_resulta}
\end{table}

\section{Conclusion}

In this article, we extend our previous line of research on suggesting MeSH terms for Boolean queries for systematic review literature search. This task adds to a recent stream of research that has focused on computational methods for the assisted creation~\cite{scells2020automatic,scells2020objective,scells2021comparison, agosti2019analysis} or refinement~\cite{wang2021mesh, scells2018generating, scells2019www, agosti2020post, alharbi2020refining} of Boolean queries for systematic review creation. 
In addition to the lexical methods we proposed previously, in this new line of work, we introduced a new set of BERT based MeSH suggestion methods.
We undertook a comprehensive evaluation and analysis of our new MeSH suggestion methods. We compared the effectiveness of the suggested MeSH terms from our new methods to both our existing lexical methods and the original queries formulated by information specialists. We found that the MeSH terms originally chosen by information specialists were often not the most effective choice and that more effective MeSH terms can be suggested automatically by our new methods.
We also found that 
using BERT methods can generally achieve higher effectiveness than the Lexical method in MeSH Term suggestion: this may be due to the fact that
BERT methods were often able to capture deeper semantic relationships. This finding motivates future work to combine lexical and BERT methods in order to reap the benefits of both approaches. Combining such sparse and dense approaches has seen much success in related areas of research, such as ad-hoc search~\cite{wang2021bert, li2022interpolate, karpukhin2020dense, ma2021replication}. 

In addition, we believe that the full potential of using MeSH entities in our suggestion method is unexplored. In our future work, we project three research directions using more information from MeSH entities to achieve more effective MeSH term suggestions, including (1) Use of MeSH tree hierarchy: MeSH entities are organised in a tree hierarchy. The parent-child relationship of entities may further restrict the number of MeSH terms suggested by MeSH Term suggestion methods (Exp: Use parent MeSH entity to restrict which child entities can appear in the suggestion list). (2) Use of MeSH categories: MeSH entities are categorised according to their natures (term, concept, descriptor and category). The nature of MeSH entities may be used in the fine-tuning process of the MeSH term suggestion methods to represent the MeSH entities. (3) Use of external MeSH definition: Each MeSH entity has a corresponding Wikipedia page to explain its content and uses. These comprehensive pages may be used to further fine-tune our MeSH term suggestion model to achieve effective MeSH term suggestions.

Identifying MeSH terms to add to a Boolean query for a systematic review literature search is a difficult task for information specialists. The findings of this article have implications for both the Information Retrieval and Systematic Review communities. Firstly, our methods can be used in automatic query formulation situations (see, e.g., work by~\citet{scells2021comparison}). Secondly, they can be integrated into existing tools to assist information specialists in formulating more effective queries~\cite{scells2018searchrefiner,li2020systematic}.






\section{Appendices}

\begin{landscape}
\begin{table}
	\centering
	\footnotesize
		\begin{tabular}{p{1pt}l|cccc|cccc|cccc|cccc}
		\\ \toprule
		\multicolumn{2}{c|}{Dataset}&\multicolumn{4}{c|}{2017}&\multicolumn{4}{c|}{2018}&\multicolumn{4}{c|}{2019-dta}&\multicolumn{4}{c}{2019-intervention}\\ \midrule
		\multicolumn{2}{c|}{Method}&\multicolumn{1}{c}{P}&\multicolumn{1}{c}{F1}&\multicolumn{1}{c}{F3}&\multicolumn{1}{c|}{R}&\multicolumn{1}{c}{P}&\multicolumn{1}{c}{F1}&\multicolumn{1}{c}{F3}&\multicolumn{1}{c|}{R}&\multicolumn{1}{c}{P}&\multicolumn{1}{c}{F1}&\multicolumn{1}{c}{F3}&\multicolumn{1}{c|}{R}&\multicolumn{1}{c}{P}&\multicolumn{1}{c}{F1}&\multicolumn{1}{c}{F3}&\multicolumn{1}{c}{R}\\ \midrule
		
		&ORIGINAL&0.0288&0.0311&0.0440&0.7745&0.0323&0.0576&0.0965&0.8629&0.0227&\textbf{0.0421}&\textbf{0.0738}&0.8966&0.0165&0.0212&0.0309&0.7471\\\midrule
		\multirow{8}{*}{\rotatebox{90}{Lexical Method}}&ATM&0.0265&0.0262&0.0353&0.7549&0.0317&0.0552&0.0898&0.8190&0.0113&0.0211&0.0373&0.8916&0.0156&0.0183&0.0269&0.7073\\
		&ATM-CUT&0.0316&0.0299&0.0404&0.7269&0.0354&0.0624&0.1033&0.7998&\textbf{0.0243}&0.0398&0.0637&0.8375&0.0173&0.0191&0.0288&0.6938\\
		&MetaMap&0.0304&0.0287&0.0381&0.7519&0.0342&0.0599&0.0980&0.8150&0.0131&0.0245&0.0433&0.8791&0.0135&0.0218&0.0339&0.6974\\
		&MetaMap-CUT&0.0337&0.0312&0.0423&0.7191&0.0360&0.0633&0.1043&0.8071&0.0193&0.0358&0.0625&0.8393&0.0159&0.0251&0.0382&0.6831\\
		&UMLS&0.0275&0.0269&0.0355&0.7458&0.0297&0.0519&0.0847&0.8200&0.0114&0.0214&0.0384&0.8616&0.0118&0.0183&0.0275&0.6998\\
		&UMLS-CUT&0.0335&0.0315&0.0430&0.7225&0.0384&\textbf{0.0681}&\textbf{0.1133}&0.7963&0.0174&0.0305&0.0508&0.8381&0.0173&0.0191&0.0295&0.6638\\
		&Fusion&0.0218&0.0227&0.0300&0.7712&0.0284&0.0495&0.0800&0.8455&0.0103&0.0192&0.0342&0.9075&0.0109&0.0173&0.0263&0.7212\\
		&Fusion-CUT&0.0323&0.0303&0.0409&0.7282&0.0333&0.0582&0.0951&0.8120&0.0147&0.0269&0.0465&0.8394&0.0161&0.0173&0.0262&0.6797\\\midrule
		\multirow{6}{*}{\rotatebox{90}{BERT Method}}&Atomic-BERT-FO&0.0257&0.0249&0.0330&\textbf{0.7830}&0.0289&0.0488&0.0795&0.8523&0.0092&0.0173&0.0310&0.8870&0.0070&0.0126&0.0219&0.7587\\
		&Semantic-BERT-FO&0.0273&0.0272&0.0363&0.7633&0.0284&0.0501&0.0820&0.8502&0.0096&0.0181&0.0324&0.8870&0.0110&0.0183&0.0288&0.7483\\
		&Fragment-BERT-FO&\textbf{0.0342}&\textbf{0.0324}&\textbf{0.0446}&0.7415&\textbf{0.0382}&0.0678&0.1132&0.8041&0.0169&0.0314&0.0548&0.8924&\textbf{0.0212}&\textbf{0.0276}&\textbf{0.0422}&0.7106\\
		&Fragment-BERT-SA&0.0212&0.0216&0.0284&0.7699&0.0268&0.0471&0.0772&\textbf{0.8652}&0.0097&0.0181&0.0323&\textbf{0.9357}&0.0076&0.0137&0.0235&\textbf{0.7806}\\
		&Fragment-BERT-SO&0.0265&0.0250&0.0335&0.7593&0.0328&0.0588&0.0991&0.8258&0.0129&0.0243&0.0433&0.8987&0.0176&0.0238&0.0358&0.7431\\
		&Fragment-BERT-LN&0.0265&0.0274&0.0373&0.7615&0.0318&0.0561&0.0925&0.8355&0.0112&0.0211&0.0378&0.8969&0.0105&0.0167&0.0265&0.7428\\
		
		\bottomrule
	\end{tabular}
	\caption{Search effectiveness of Boolean query using suggested MeSH terms evaluated by precision (P), F1,  F3 and recall (R). Lexical methods: For each method, \textit{CUT} indicates cut-off ranks.  BERT methods: \textit{FO}, \textit{SA}, \textit{SO}, \textit{LN} indicate different cut-off strategies. No statistical significant differences are detected between the ORIGINAL query and those obtained by the other methods (two-tailed t-test with Bonferroni correction, $p<0.05$).}	
	\vspace{-10pt}
	\label{table:search_result}
	
\end{table}

\begin{table}
	\centering
	\footnotesize
	\begin{tabular}{p{1pt}l|cccc|cccc|cccc|cccc}
		\\ \toprule
		\multicolumn{2}{c|}{Dataset}&\multicolumn{4}{c|}{2017}&\multicolumn{4}{c|}{2018}&\multicolumn{4}{c|}{2019-dta}&\multicolumn{4}{c}{2019-intervention}\\ \midrule
		\multicolumn{2}{c|}{Method}&\multicolumn{1}{c}{P}&\multicolumn{1}{c}{F1}&\multicolumn{1}{c}{F3}&\multicolumn{1}{c|}{R}&\multicolumn{1}{c}{P}&\multicolumn{1}{c}{F1}&\multicolumn{1}{c}{F3}&\multicolumn{1}{c|}{R}&\multicolumn{1}{c}{P}&\multicolumn{1}{c}{F1}&\multicolumn{1}{c}{F3}&\multicolumn{1}{c|}{R}&\multicolumn{1}{c}{P}&\multicolumn{1}{c}{F1}&\multicolumn{1}{c}{F3}&\multicolumn{1}{c}{R}\\ \midrule
		\multirow{8}{*}{\rotatebox{90}{Lexical Method}}&ATM&0.9148&0.7515&0.6465&0.8570&0.9670&0.9216&0.8578&0.5820&0.4586&0.4573&0.4544&0.9674&0.9400&0.7730&0.7292&0.7467\\
		
		&ATM-CUT&0.8992&0.9396&0.8480&0.6698&0.8426&0.8532&0.4265&0.9196&0.9378&0.8354&0.6629&0.9514&0.8275&0.8556&0.6688&0.8306\\
		&MetaMap&0.9411&0.8784&0.7572&0.8366&0.9263&0.9668&0.5474&0.5297&0.5302&0.5294&0.8930&0.7702&0.9612&0.8361&0.6867&0.8956\\
		&MetaMap-CUT&0.8279&0.9954&0.9320&0.6203&0.8159&0.8330&0.4813&0.8305&0.8271&0.8205&0.6691&0.9502&0.7414&0.6170&0.6078&0.7990\\
		&UMLS&0.9468&0.7873&0.6529&0.7933&0.8132&0.7511&0.5876&0.4432&0.4475&0.4535&0.7712&0.6556&0.8076&0.8162&0.7007&0.8556\\
		&UMLS-CUT&0.8323&0.9785&0.9583&0.6363&0.6625&0.6500&0.4027&0.7251&0.6698&0.6245&0.6667&0.9508&0.8238&0.8980&0.5085&0.6661\\
		&Fusion&0.7208&0.5824&0.4505&0.9755&0.7373&0.6598&0.8258&0.4014&0.4031&0.4049&0.9199&0.5801&0.7308&0.7487&0.8293&0.7838\\
		&Fusion-CUT&0.8768&0.9598&0.8729&0.6786&0.9817&0.9714&0.5223&0.5972&0.5856&0.5704&0.6722&0.9726&0.6744&0.6682&0.5919&0.9463\\\midrule
		\multirow{6}{*}{\rotatebox{90}{BERT Method}}&Atomic-BERT-FO&0.8905&0.7021&0.5662&0.9363&0.8109&0.7149&0.6493&0.8917&0.3493&0.3526&0.3566&0.9410&0.2777&0.3441&0.4487&0.9215\\
		&Semantic-BERT-FO&0.9465&0.8073&0.6916&0.9164&0.7530&0.6962&0.8704&0.3615&0.3657&0.3711&0.9410&0.5684&0.7928&0.8853&0.9918&0.7841\\
		&Fragment-BERT-FO&0.8075&0.9371&0.9712&0.7663&0.6737&0.6540&0.4589&0.7074&0.7015&0.6948&0.9715&0.7003&0.5612&0.4209&0.7636&0.6814\\
		&Fragment-BERT-SA&0.7171&0.5460&0.3968&0.9658&0.6480&0.5888&0.9754&0.3723&0.3748&0.3777&0.6593&0.3085&0.4157&0.5481&0.7687&0.6837\\
		&Fragment-BERT-SO&0.9135&0.6889&0.5615&0.8908&0.9592&0.9406&0.6374&0.5109&0.5154&0.5208&0.9856&0.9295&0.8150&0.7191&0.9736&0.9739\\
		&Fragment-BERT-LN&0.9119&0.8152&0.7293&0.9053&0.9452&0.9110&0.7120&0.4368&0.4416&0.4474&0.9978&0.5192&0.6417&0.7200&0.9715&0.9681\\
		
		\bottomrule
	\end{tabular}
	\caption{ Two-tailed t-test results of Boolean query search effectiveness  between the ORIGINAL query and those obtained by the other methods by precision (P), F1,  F3 and recall (R). Lexical methods: \textit{CUT} indicates cut-off ranks.  BERT methods: \textit{FO}, \textit{SA}, \textit{SO}, \textit{LN} indicate different cut-off strategies.}	
	\vspace{-10pt}
	\label{table:search_result_p}
	
\end{table}
\end{landscape}

\begin{landscape}
\begin{table*}
	\centering
	\scriptsize
	\begin{tabular}{l|p{175pt}|c|p{335pt}}
		\multicolumn{4}{c}{}\\	\toprule
		Topic&\multicolumn{3}{c}{CD009642}\\\midrule
		Fragments&\multicolumn{1}{c}{Fragment 1}&\multicolumn{1}{c}{}&\multicolumn{1}{c}{Fragment 2} \\\midrule
		ORIGINAL& \textbf{Lidocaine} OR lidocain* OR  Lignocain* OR Xylocain* &&\textbf{Pain} OR \textbf{Pain, Postoperative} OR \textbf{Postoperative Care} OR \textbf{Postoperative Complications} OR (post operative OR postoperative) AND (pain* OR recovery) \\\cmidrule{1-2}\cmidrule{4-4}
		ATM & lidocain* OR Lignocain* OR Xylocain*&& \textbf{Pain} OR (post operative OR postoperative) AND (pain* OR recovery) \\\cmidrule{1-2}\cmidrule{4-4}
		ATM-CUT &  lidocain* OR Lignocain* OR Xylocain* &&\textbf{Pain} OR (post operative OR postoperative) AND (pain* OR recovery)\\\cmidrule{1-2}\cmidrule{4-4}
		MetaMap& \textbf{Lidocaine} OR lidocain* OR Lignocain* OR Xylocain*&& \textbf{Pain} OR (post operative OR postoperative) AND (pain* OR recovery) \\\cmidrule{1-2}\cmidrule{4-4}
		MetaMap-CUT &\textbf{Lidocaine} OR lidocain* OR Lignocain* OR Xylocain*  &&\textbf{Pain} OR (post operative OR postoperative) AND (pain* OR recovery)\\\cmidrule{1-2}\cmidrule{4-4}
		UMLS& \textbf{Lidocaine} OR lidocain* OR Lignocain* OR Xylocain*&& \textbf{Pain} OR (post operative OR postoperative) AND (pain* OR recovery) \\\cmidrule{1-2}\cmidrule{4-4}
		UMLS-CUT & \textbf{Lidocaine} OR lidocain* OR Lignocain* OR Xylocain* && \textbf{Pain} OR (post operative OR postoperative) AND (pain* OR recovery) \\\cmidrule{1-2}\cmidrule{4-4}
		Fusion & \textbf{Lidocaine} OR lidocain* OR Lignocain* OR Xylocain*&\multirow{6}{*}{AND}& \textbf{Pain} OR (post operative OR postoperative) AND (pain* OR recovery) \\\cmidrule{1-2}\cmidrule{4-4}
		Fusion-CUT & \textbf{Lidocaine} OR lidocain* OR Lignocain* OR Xylocain*&& \textbf{Pain} OR (post operative OR postoperative) AND (pain* OR recovery) \\\cmidrule{1-2}\cmidrule{4-4}
		Atomic-BERT-FO & \textbf{Lidocaine} OR \textbf{Xylans} OR lidocain* OR Lignocain* OR Xylocain*&& \textbf{Postoperative Care} OR \textbf{Pain} OR \textbf{Recovery of Function} OR (post operative OR postoperative) AND (pain* OR recovery) \\\cmidrule{1-2}\cmidrule{4-4}
		Semantic-BERT-FO & \textbf{Lidocaine} OR \textbf{Xylans} OR lidocain* OR Lignocain* OR Xylocain*&& \textbf{Postoperative Care} OR \textbf{Pain} OR \textbf{Recovery of Function} OR (post operative OR postoperative) AND (pain* OR recovery) \\\cmidrule{1-2}\cmidrule{4-4}
		Fragment-BERT-FO & \textbf{Lidocaine} OR lidocain* OR Lignocain* OR Xylocain*&& \textbf{Pain, Postoperative} OR (post operative OR postoperative) AND (pain* OR recovery) \\\cmidrule{1-2}\cmidrule{4-4}
		Fragment-BERT-SA & \textbf{Lidocaine} OR \textbf{Procaine} OR \textbf{Xylans} OR lidocain* OR Lignocain* OR Xylocain*&& \textbf{Pain, Postoperative} OR \textbf{Postoperative Care} OR \textbf{Postoperative Period} OR \textbf{Postoperative Complications} OR (post operative OR postoperative) AND (pain* OR recovery) \\\cmidrule{1-2}\cmidrule{4-4}
		Fragment-BERT-SO & \textbf{Lidocaine} OR lidocain* OR Lignocain* OR Xylocain*&& \textbf{Pain, Postoperative} OR \textbf{Postoperative Care} OR \textbf{Postoperative Period} OR \textbf{Postoperative Complications} OR (post operative OR postoperative) AND (pain* OR recovery) \\\cmidrule{1-2}\cmidrule{4-4}
		Fragment-BERT-LN & \textbf{Lidocaine} OR \textbf{Procaine} OR \textbf{Xylans} OR lidocain* OR Lignocain* OR Xylocain*&& \textbf{Pain, Postoperative} OR \textbf{Postoperative Care} OR \textbf{Postoperative Period} OR (post operative OR postoperative) AND (pain* OR recovery) \\
		\bottomrule	\end{tabular}
	\caption{Query fragments in different methods, For Lexical methods: \textit{CUT} indicates cut-off ranks . For BERT suggestion method, \textit{A-B}indicates \atb, \textit{S-B} indicates \seb, \textit{F-B} indicates \frb. In each BERT method, \textit{FO}, \textit{SA}, \textit{SO}, \textit{LN} indicates cut-off strategy used. For each fragment, bold text means MeSH term.}	
	\label{table:casestudy_better}
\end{table*}
\end{landscape}
\begin{landscape}
\begin{table*}
	\centering
	\scriptsize
	\begin{tabular}{l|p{120pt}|c|p{360pt}}
		\multicolumn{4}{c}{}\\ \toprule
		Topic& \multicolumn{3}{c}{CD004414}\\\midrule
		Fragments&\multicolumn{1}{c}{Fragment 1} &\multicolumn{1}{c}{}&\multicolumn{1}{c}{Fragment 2}\\\midrule
		ORIGINAL& \textbf{Hand} OR hand* OR finger* OR palm*&& \textbf{Hand Dermatoses} OR (dermat* OR eczema) AND (occupation* OR irritant* OR contact) AND (hand* OR finger* OR palm*) \\\cmidrule{1-2}\cmidrule{4-4}
		ATM & \textbf{Hand} OR \textbf{Fingers} OR hand* OR finger* OR palm* && \textbf{Fingers} OR \textbf{Eczema} OR \textbf{Hand} OR \textbf{Irritants} OR \textbf{Occupations} OR (dermat* OR eczema) AND (occupation* OR irritant* OR contact) AND (hand* OR finger* OR palm*)\\\cmidrule{1-2}\cmidrule{4-4}
		ATM-CUT & \textbf{Hand} OR hand* OR finger* OR palm* && \textbf{Fingers} OR \textbf{Eczema} OR (dermat* OR eczema) AND (occupation* OR irritant* OR contact) AND (hand* OR finger* OR palm*)\\\cmidrule{1-2}\cmidrule{4-4}
		MetaMap & \textbf{Hand} OR \textbf{Fingers} OR hand* OR finger* OR palm* && \textbf{Hand} OR \textbf{Fingers} OR \textbf{Eczema} OR \textbf{Occupations} OR \textbf{Irritants} OR (dermat* OR eczema) AND (occupation* OR irritant* OR contact) AND (hand* OR finger* OR palm*)\\\cmidrule{1-2}\cmidrule{4-4}
		MetaMap-CUT & \textbf{Hand} OR hand* OR finger* OR palm* && \textbf{Hand} OR \textbf{Fingers} OR \textbf{Eczema} OR (dermat* OR eczema) AND (occupation* OR irritant* OR contact) AND (hand* OR finger* OR palm*)\\\cmidrule{1-2}\cmidrule{4-4}
		UMLS & \textbf{Fingers} OR \textbf{Hand} OR \textbf{Palm Oil} OR \textbf{Computers, Handheld} OR hand* OR finger* OR palm* && \textbf{Eczema} OR \textbf{Fingers} OR \textbf{Hand} OR \textbf{Dermatitis, Atopic} OR \textbf{Kaposi Varicelliform Eruption} OR \textbf{Retirement} OR \textbf{Computers, Handheld} OR \textbf{Occupations} OR \textbf{Palm Oil} OR \textbf{Irritants} OR (dermat* OR eczema) AND (occupation* OR irritant* OR contact) AND (hand* OR finger* OR palm*)\\\cmidrule{1-2}\cmidrule{4-4}
		UMLS-CUT & \textbf{Fingers} OR hand* OR finger* OR palm* && \textbf{Eczema} OR \textbf{Fingers} OR (dermat* OR eczema) AND (occupation* OR irritant* OR contact) AND (hand* OR finger* OR palm*)\\\cmidrule{1-2}\cmidrule{4-4}
		Fusion & \textbf{Hand} OR \textbf{Fingers} OR \textbf{Palm Oil} OR \textbf{Computers, Handheld} OR hand* OR finger* OR palm* &\multirow{6}{*}{AND}& \textbf{Eczema} OR \textbf{Fingers} OR \textbf{Hand} OR \textbf{Dermatitis, Atopic} OR \textbf{Occupations} OR \textbf{Kaposi Varicelliform Eruption} OR \textbf{Retirement} OR \textbf{Irritants} OR \textbf{Computers, Handheld} OR \textbf{Palm Oil} OR (dermat* OR eczema) AND (occupation* OR irritant* OR contact) AND (hand* OR finger* OR palm*)\\\cmidrule{1-2}\cmidrule{4-4}
		Fusion-CUT & \textbf{Hand} OR hand* OR finger* OR palm* && \textbf{Eczema} OR \textbf{Fingers} OR \textbf{Hand} OR (dermat* OR eczema) AND (occupation* OR irritant* OR contact) AND (hand* OR finger* OR palm*)\\\cmidrule{1-2}\cmidrule{4-4}
		Atomic-BERT-FO & \textbf{Hand} OR \textbf{Fingers} OR \textbf{Palm Oil} OR hand* OR finger* OR palm* && \textbf{Hand} OR \textbf{Eczema} OR \textbf{Occupations} OR \textbf{Dermatology} OR \textbf{Irritants} OR \textbf{Dermatitis, Contact} OR \textbf{Fingers} OR \textbf{Palm Oil} OR (dermat* OR eczema) AND (occupation* OR irritant* OR contact) AND (hand* OR finger* OR palm*)\\\cmidrule{1-2}\cmidrule{4-4}
		Semantic-BERT-FO & \textbf{Hand} OR \textbf{Fingers} OR \textbf{Palm Oil} OR hand* OR finger* OR palm* && \textbf{Dermatology} OR \textbf{Eczema} OR \textbf{Occupations} OR \textbf{Irritants} OR \textbf{Dermatitis, Contact} OR \textbf{Hand} OR \textbf{Fingers} OR \textbf{Palm Oil} OR (dermat* OR eczema) AND (occupation* OR irritant* OR contact) AND (hand* OR finger* OR palm*)\\\cmidrule{1-2}\cmidrule{4-4}
		Fragment-BERT-FO & \textbf{Hand} OR hand* OR finger* OR palm* && \textbf{Dermatitis, Contact} OR (dermat* OR eczema) AND (occupation* OR irritant* OR contact) AND (hand* OR finger* OR palm*)\\\cmidrule{1-2}\cmidrule{4-4}
		Fragment-BERT-SA & \textbf{Hand} OR \textbf{Fingers} OR \textbf{Palm Oil} OR hand* OR finger* OR palm* && \textbf{Dermatitis, Contact} OR \textbf{Dermatitis, Allergic Contact} OR \textbf{Hand} OR \textbf{Fingers} OR \textbf{Eczema} OR \textbf{Dermatitis, Atopic} OR \textbf{Patch Tests} OR \textbf{Skin Diseases} OR (dermat* OR eczema) AND (occupation* OR irritant* OR contact) AND (hand* OR finger* OR palm*)\\\cmidrule{1-2}\cmidrule{4-4}
		Fragment-BERT-SO & \textbf{Hand} OR hand* OR finger* OR palm* && \textbf{Dermatitis, Contact} OR (dermat* OR eczema) AND (occupation* OR irritant* OR contact) AND (hand* OR finger* OR palm*)\\\cmidrule{1-2}\cmidrule{4-4}
		Fragment-BERT-LN & \textbf{Hand} OR \textbf{Fingers} OR \textbf{Palm Oil} OR hand* OR finger* OR palm* && \textbf{Dermatitis, Contact} OR \textbf{Dermatitis, Allergic Contact} OR \textbf{Hand} OR (dermat* OR eczema) AND (occupation* OR irritant* OR contact) AND (hand* OR finger* OR palm*)\\
		
		\bottomrule	\end{tabular}
	\caption{Query fragments in different methods, For Lexical methods: \textit{CUT} indicates cut-off ranks . For BERT suggestion method, \textit{A-B}indicates \atb, \textit{S-B} indicates \seb, \textit{F-B} indicates \frb. In each BERT method, \textit{FO}, \textit{SA}, \textit{SO}, \textit{LN} indicates cut-off strategy used. For each fragment, bold text means MeSH term.}	
	\label{table:casestudy_worse}
\end{table*}

\end{landscape}

\section{Acknowledgments}
Shuai Wang is supported by a UQ Earmarked PhD Scholarship and this research is funded by the Australian Research Council Discovery Projects program ARC Discovery Project DP210104043.





\bibliographystyle{elsarticle-num-names}
\bibliography{references.bib}







\end{document}